\documentclass[12pt]{JHEP3}

\usepackage{graphicx}

\usepackage{amsmath}

\newcommand{\bea}{\begin{eqnarray}}
\newcommand{\eea}{\end{eqnarray}}

\makeatletter
\@addtoreset{equation}{section}
\makeatother

\title{
Universal turbulence on branes in holography
}
\author{
Koji Hashimoto$^{*}$, 
Mitsuhiro Nishida$^{\#}$ and Akihiko Sonoda$^{\dagger}$
\\

{\it Department of Physics, Osaka University,
Toyonaka, Osaka 560-0043, Japan}\\
E-mail: $^*$ \email{koji(at)phys.sci.osaka-u.ac.jp}\\ 
E-mail: $^\#$ \email{nishida(at)het.phys.sci.osaka-u.ac.jp}\\
E-mail: $^\dagger$ \email{sonoda(at)het.phys.sci.osaka-u.ac.jp}\\ 
}

\abstract{At a meson melting transition in holographic QCD,
a weak turbulence of mesons was found 
with critical embeddings of probe D-branes in gravity duals.
The turbulent mesons have a power-law energy distribution 
$\varepsilon_n \propto (\omega_n)^\alpha$ where $\omega_n$ is the mass
of the $n$-th excited resonance of the meson tower. In this paper, we find that
the turbulence power $\alpha$ is universal, irrespective of how the transition is
driven, by numerically calculating the power in various static brane setups at criticality. 
We also find that the power
$\alpha$ depends only on the cone dimensions of the probe D-branes.
}

\preprint{OU-HET 860}

\begin{document}
\maketitle
\setcounter{page}{1}

\section{Turbulence on branes in holography}
\label{sec1}

AdS instability conjecture by 
Bizon and Rostworowski \cite{Bizon:2011gg} has attracted much attention on
intrinsic turbulent nature of AdS spacetime. The importance of addressing 
the stability question of generic AdS spacetimes is obvious from the viewpoint of
the renowned AdS/CFT correspondence
\cite{Maldacena:1997re,Gubser:1998bc,Witten:1998qj}. 

The issue of the AdS instability 
brings about various fruitful discussions on a possible 
universality of the instability
(see for example \cite{deOliveira:2012dt}-\cite{Wu:2012rib}),
while 
partially relies on details of numerical simulations.
On the other hand, probe D-branes in AdS spacetimes, which are quite popular in
AdS/CFT correspondence for introduction of quarks \cite{Karch:2002sh,Kruczenski:2003be,Kruczenski:2003uq,Sakai:2004cn,Erdmenger:2007cm} and on which AdS spacetimes are generically induced, may provide
an easier toy model for a similar turbulent instability. In fact, 
in Refs.\cite{Hashimoto:2014xta,Hashimoto:2014dda}, turbulent spectra of meson energy are observed, which is quite reminiscent
of the gravity instability in AdS spacetimes. 
The scalar and vector fields on the flavor D7-brane in $AdS_5\times S^5$ geometry
correspond to mesons in ${\cal N}=2$ supersymmetric QCD 
{\it a la} AdS/CFT, and a decomposition by the discrete
momentum along the holographic (AdS radial) direction defines 
the energy distribution for the $n$-th level of meson excitation. It was found in
\cite{Hashimoto:2014xta}\cite{Hashimoto:2014dda} 
that the energy distribution at a critical brane embedding forming a conical shape
with the help of background electric field exhibits
\begin{eqnarray}
\varepsilon_n \propto (\omega_n)^{-5} \, .
\label{turbulent}
\end{eqnarray}
Here $\varepsilon_n$ is the energy of the $n$-th resonance of the mesons,
and $\omega_n$ is the mass of the $n$-th resonant meson. Since the mass
$\omega_n$ is nothing but the discrete momentum along the holographic direction,
the critical behavior (\ref{turbulent}) is a weak turbulence.

The interesting points in \cite{Hashimoto:2014xta}\cite{Hashimoto:2014dda}
about the brane turbulence are that the turbulence (\ref{turbulent}) appears
not only in time-dependent simulations of applied electric fields 
\cite{Hashimoto:2014yza, Ishii:2014paa}
but also for a static brane configuration under a static external electric field, 
which is called critical embedding \cite{Erdmenger:2007bn,Albash:2007bq}
 (see \cite{Mateos:2006nu} and
\cite{Frolov:2006tc} for the critical embedding
at thermal phase transition). This is the reason why the brane
turbulence may be tractable in a easier manner.

In this paper, we analyze the weak turbulence on the probe D-branes in AdS
geometries, and find a universality of the turbulent behavior, and also
a universality of the exponent in the relation 
\begin{eqnarray}
\varepsilon_n \propto (\omega_n)^{\alpha} \, .
\label{turbulent2}
\end{eqnarray}
Turning on electric fields and magnetic fields, and also temperature,
we find that this power $\alpha$ does not change at criticality. 
It is dependent only
on dimensions of the probe brane.

Let us describe a little more on what is the criticality where the turbulent
behavior (\ref{turbulent2}) appears.
The quark-hadron transition in strongly coupled gauge theories including QCD is one
of the famous transitions, and can be realized either by turning on a
nonzero temperature, or by putting a strong electric field, or by some other methods. 
The phase boundary separates the hadron (``confining") phase and 
the quark (``deconfining") phase. In large $N_c$ gauge theories where $N_c$ is the
number of colors, the phase transition can be first order. 
In the AdS/CFT analysis in a probe approximation of a flavor brane, 
there appears a first order phase transition called meson melting \cite{Mateos:2006nu}.
When the temperature, or the external electric field applied to the quarks, 
is not large, the theory has a discrete spectrum of mesons. On the other hand,
when it is large, the meson spectrum becomes continuous, and mesons melt.
The transition shares properties with the quark-hadron phase transition in QCD
(although generically gluons are always deconfined in this meson melting transition).

The meson melting transition has been known to be accompanied by 
a conic D-brane configuration (the critical embedding). 
It appears at the phase boundary,
and should characterize the meson melting transition.
The conic shape of the probe D-brane can be rephrased in terms of 
condensation of mesons, as the position of the D-brane in the bulk gravity geometry
is probed by the expectation value of the mesons. 
In fact, for the case of a D7-brane in static electric fields, the critical embedding
shows \cite{Hashimoto:2014xta,Hashimoto:2014dda}
the turbulent condensation of mesons characterized by
(\ref{turbulent}). The relation (\ref{turbulent}) was also kept for dynamical
simulation for time-dependent electric field backgrounds or time-dependent
quark mass backgrounds \cite{Hashimoto:2014xta,Hashimoto:2014dda}.

In this paper, we examine the relation (\ref{turbulent2}) for various static 
backgrounds providing critical embedding of probe D-branes. In Section 2,
we consider generic electric and magnetic field on probe D7-branes in
AdS, and also a temperature, and find a universal power 
\begin{eqnarray}
\alpha = -5  \qquad (\mbox{D7 in $AdS_5$}).
\end{eqnarray}
In Section 3, we examine the case of a probe D5-brane in similar external fields,
and find a universal power 
\begin{eqnarray}
\alpha = -4  \qquad (\mbox{D5 in $AdS_5$}).
\end{eqnarray}
From these results and also some simple analysis of conic branes of 1 and 2 dimensions in flat space, in Section 4 we make a conjecture for generic
brane turbulence
\begin{eqnarray}
\varepsilon_n \propto (\omega_n)^{-d_{\rm cone}-1} \, ,
\end{eqnarray}
where $d_{\rm cone}$ is the number of the spatial dimensions of the cone
of the brane.

\section{D3-D7 system and universal turbulence}

In this section, we analyze the meson turbulence in the D3-D7 system with an electromagnetic field or at a finite temperature, by using the AdS/CFT correspondence. 
The turbulent behavior was found in this probe D7-brane system in \cite{Hashimoto:2014xta}, and here 
we explore different parameter space as external fields, to find that the turbulence
power spectrum found in \cite{Hashimoto:2014xta} is universal: $\alpha = -5$.

First, we review the D3-D7 system and the scalar meson mass spectrum. Next, we analyze a relation between the energy of the meson $\varepsilon_n$ and the meson mass $\omega_n$ by changing the electromagnetic field and the temperature. We will find that
the universal power-law of the energy of the higher states of the meson resonances 
goes as $\varepsilon_n\propto\omega_n^{-5}$, when the probe D-brane is 
at the critical embedding. 

\subsection{Review of the $\mathcal{N}=2$ supersymmetric QCD in AdS/CFT}

First, to fix our notation, we briefly review the effective action of mesons for the four-dimensional $\mathcal{N}=2$ supersymmetric QCD given by a gravity dual of the D3-D7 system \cite{Karch:2002sh}. In order to use the AdS/CFT correspondence, we consider the gauge group $U(N_c)$ and one-flavor quark sector with a large $N_c$ limit $N_c\to\infty$ and the strong coupling limit $\lambda\equiv N_cg^2_{\textrm{YM}}\to\infty$. The brane configuration of the D3-D7 system is shown in the following table:
\\

\hspace{20mm}
\begin{tabular}{|c|c|c|c|c|c|c|c|c|c|c|} 
\hline
      & 0 & 1 & 2 & 3 & 4 & 5 & 6 & 7 & 8 & 9 \\ \hline
D3 & $\surd$ & $\surd$ &  $\surd$ & $\surd$ &  &  &  &  &  &   \\ \hline
D7 & $\surd$ & $\surd$ & $\surd$ & $\surd$ & $\surd$ & $\surd$ & $\surd$ & $\surd$ &\hspace{3mm} & \hspace{3mm}  \\ 
\hline 
\end{tabular}
\\ 

In the gravity dual, the effective action of the meson tower is
the flavor D7-brane action in the $\textrm{AdS}_5\times S^5$ background geometry, 
\begin{align}
&S=\frac{-1}{(2\pi)^6g_{\textrm{YM}}^2l_s^8}\int d^8 \xi \sqrt{-\det(g_{ab}[w]+2\pi l_s^2F_{ab})},\label{DBI D3-D7}\\
&ds^2=\frac{r^2}{R^2}\eta_{\mu\nu}dx^\mu dx^\nu+\frac{R^2}{r^2}\left[d\rho^2+\rho^2d\Omega_3^2+dw^2+d\bar{w}^2\right],
\end{align}
where $r^2\equiv\rho^2+w^2+\bar{w}^2, F_{ab}=\partial_aA_b-\partial_bA_a,$ and $R\equiv(2\lambda)^{1/4}l_s$ is the $\textrm{AdS}_5$ radius. We treat the flavor D7-brane as a probe, so it does not affect the $\textrm{AdS}_5\times S^5$ background geometry, which is justified at $N_c\gg1$. For our purpose we can assume that the fields of the action depend only on $x^\mu$ and $\rho$ such that $w=w(x^\mu,\rho)$, $\bar{w}=\bar{w}(x^\mu, \rho)$ and $A_a = A_a(x^\mu, \rho)$. These fields correspond to the scalar mesons and the vector mesons, after they are decomposed as shown below. 

In order to solve the equation of the motion for (\ref{DBI D3-D7}), a boundary condition 
\begin{align}
w(x^\mu, \rho=\infty)=R^2m,\label{bc D3-D7}
\end{align} 
is necessary (while we set $\bar{w}(x^\mu, \rho)=0$ by using a rotation symmetry of the $(w, \bar{w})$-plane). 
In the AdS/CFT correspondence, $m$ is related to the quark mass $m_q$ as $m_q=(\lambda /2\pi^2)^{1/2}m$. The static solution of the equation of the motion for (\ref{DBI D3-D7}) with (\ref{bc D3-D7}) is 
\begin{align}
w(x^\mu, \rho)=R^2m, \;\;\;\;\;A_a(x^\mu, \rho)=0,\label{static solution}
\end{align}
which is a straight D7-brane located at $w(x^\mu, \rho)=R^2m$.

From now, we consider a scalar fluctuation of $w(x^\mu, \rho)$ around the solution (\ref{static solution}) \cite{Kruczenski:2003be}. We define the scalar fluctuation field as $\chi(t,\rho)\equiv R^{-2}w-m$ where for simplicity we assume that it does not depend on $x^1, x^2, x^3.$ Substituting it to (\ref{DBI D3-D7}), we obtain an action of $\chi$ as
\begin{align}
S=\int dt d^3 x \int^\infty_0 d\rho\frac{\rho^3}{2(\rho^2+R^4m^2)^2}\left[(\partial_t\chi)^2-\frac{(\rho^2+R^4m^2)^2}{R^4}(\partial_\rho\chi)^2\right]+\mathcal{O}(\chi^3),\label{chiaction}
\end{align}
up to an overall factor.

In order to determine the meson mass spectrum, we solve 
the equation of the motion for $\chi$ to the first order, 
\begin{align}
\left[\frac{\partial^2}{\partial t^2}-\frac{(\rho^2+R^4m^2)^2}{\rho^3}\frac{\partial}{\partial \rho}\frac{\rho^3}{R^4}\frac{\partial}{\partial \rho}\right]\chi=0.\label{eomchi}
\end{align}
A normalizable solution of (\ref{eomchi}) is
\begin{align}
\chi&=\sum^\infty_{n=0}\textrm{Re}[C_n\exp[i\omega_nt]e_n(\rho)],\\
e_n(\rho)&\equiv\sqrt{2(2n+3)(n+1)(n+2)}\left(\frac{R^4m^2}{\rho^2+R^4m^2}\right)^{n+1}F(-n-1, -n, 2; -\frac{\rho^2}{R^4m^2}),\\
\omega_n&\equiv2\sqrt{(n+1)(n+2)}m,
\end{align}
where $F$ is the Gaussian hypergeometric function. The eigen value
$\omega_n$ corresponds to the meson mass at the resonance level $n$.

For the eigenmode expansion, we define the following inner product in the $\rho$-space,
\begin{align}
(f, g)\equiv\int^\infty_0 d\rho\frac{\rho^3}{(\rho^2+R^4m^2)^2}f(\rho)g(\rho).\label{inner product}
\end{align}
Under the inner product (\ref{inner product}), $e_n(\rho)$ forms an orthonormal basis as
\begin{align}
(e_n, e_m)=\delta_{nm}.\label{D3-D7ip}
\end{align}
By using this basis, we expand the scalar fluctuation field $\chi$ as
\begin{align}
\chi=\sum^\infty_{n=0}c_n(t)e_n(\rho).\label{chi}
\end{align}
The expansion coefficient $c_n(t)$ corresponds to the meson field 
at the resonance level $n$.

Substituting (\ref{chi}) to (\ref{chiaction}), we obtain the effective action of the meson field $c_n(t)$ as
\begin{align}
S=\frac{1}{2}\int d^4x \sum^\infty_{n=0}[\dot{c}_n^2-\omega_n^2c_n^2]+\textrm{interaction},
\end{align}
where we omit total derivative terms. We define a linearlized 
energy of the $n$-th meson based on this action as
\begin{align}
\varepsilon_n\equiv \frac{1}{2}(\dot{c}_n^2+\omega_n^2c_n^2),\label{D3-D7en}
\end{align}
and the linearized total energy as 
\begin{align}
\varepsilon\equiv\sum^\infty_{n=0}\varepsilon_n.\label{D3-D7totalen}
\end{align}
Now, we are ready for the analysis of the meson turbulence. We shall see how the total energy is distributed to the $n$-th energies $\varepsilon_n$. Irrespective of the external fields, the power law $\varepsilon_n \propto \omega_n^{-5}$ at the criticality 
will be found below.

\subsection{Turbulence with an electromagnetic field}
\label{sec:section2.2}

The turbulent meson condensation found in Ref.~\cite{Hashimoto:2014dda, Hashimoto:2014xta} at static D7-brane configuration is about a critical electric field,
and the power $\alpha$ appearing in the scaling behavior $\varepsilon_n \propto(\omega_n)^\alpha$ is found to be $\alpha = -5$. Here in this section, we
include a constant magnetic field in addition, and will find that again the
power $\alpha$ is $-5$ at criticality.

We introduce a generic constant electromagnetic field in the 
four-dimensional $x^\mu$-spacetime and calculate the power-law scaling 
of the energy of the highly excited meson states at the critical embedding. 
We turn on generic constant electromagnetic fields as
\begin{align}
&F_{01}=-F_{10}\equiv E_x, \;F_{02}=-F_{20}\equiv E_y, \;F_{03}=-F_{30}\equiv E_z,\\
&F_{12}=-F_{21}\equiv B_z, \;F_{23}=-F_{32}\equiv B_x, \;F_{31}=-F_{13}\equiv B_y,
\end{align}
and again consider a symmetric D7-brane configuration $w=w(\rho), \;\bar{w}=0$.
With this electromagnetic field, we can calculate the flavor D7-brane action as
\begin{align}
S=\int dtd^3x\int^\infty_0 d\rho\rho^3\sqrt{1+w'^2}\sqrt{1-\frac{R^4(2\pi l_s^2)^2}{(\rho^2+w^2)^2}(|\vec{E}|^2-|\vec{B}|^2)-\frac{R^8(2\pi l_s^2)^4}{(\rho^2+w^2)^4}(\vec{E}\cdot \vec{B})^2}.
\end{align}
The Euler-Lagrange equation for the brane embedding $w(\rho)$ is 
\begin{eqnarray}
0&=&\partial_\rho\left(\frac{\rho^3w'\sqrt{1-\frac{R^4(2\pi l_s^2)^2}{(\rho^2+w^2)^2}(|\vec{E}|^2-|\vec{B}|^2)-\frac{R^8(2\pi l_s^2)^4}{(\rho^2+w^2)^4}(\vec{E}\cdot \vec{B})^2}}{\sqrt{1+w'^2}}\right)\notag\\
&&
-\frac{\rho^3\sqrt{1+w'^2}\left(\frac{2R^4(2\pi l_s^2)^2w}{(\rho^2+w^2)^3}(|\vec{E}|^2-|\vec{B}|^2)+\frac{4R^8(2\pi l_s^2)^4w}{(\rho^2+w^2)^5}(\vec{E}\cdot \vec{B})^2\right)}{\sqrt{1-\frac{R^4(2\pi l_s^2)^2}{(\rho^2+w^2)^2}(|\vec{E}|^2-|\vec{B}|^2)-\frac{R^8(2\pi l_s^2)^4}{(\rho^2+w^2)^4}(\vec{E}\cdot \vec{B})^2}}.\label{D3-D7eqwithem}
\end{eqnarray}
We calculate the energy distribution $\varepsilon_n$ 
of the meson resonances by the following steps:
\begin{enumerate}
\item Solve (\ref{D3-D7eqwithem}) with (\ref{bc D3-D7}) and calculate $\chi(\rho)$ numerically.
\item By using (\ref{D3-D7ip}) and (\ref{chi}), calculate $c_n$ from $\chi(\rho)$.
\item By using the definition (\ref{D3-D7en}), calculate $\varepsilon_n$ from $c_n$.
\end{enumerate}
For numerical calculations, we set $R=m=1$, and we normalize $\mathcal{E}_i\equiv2\pi l_s^2E_i$ and $\mathcal{B}_i\equiv2\pi l_s^2B_i$. In particular, we are interested in a relation between $\varepsilon_n$ and $\omega_n$ at the critical embedding of the probe D7-brane.

\begin{figure}[thpb]
  \begin{center}
    \includegraphics[width=7cm,bb=0 0 360 232 ]{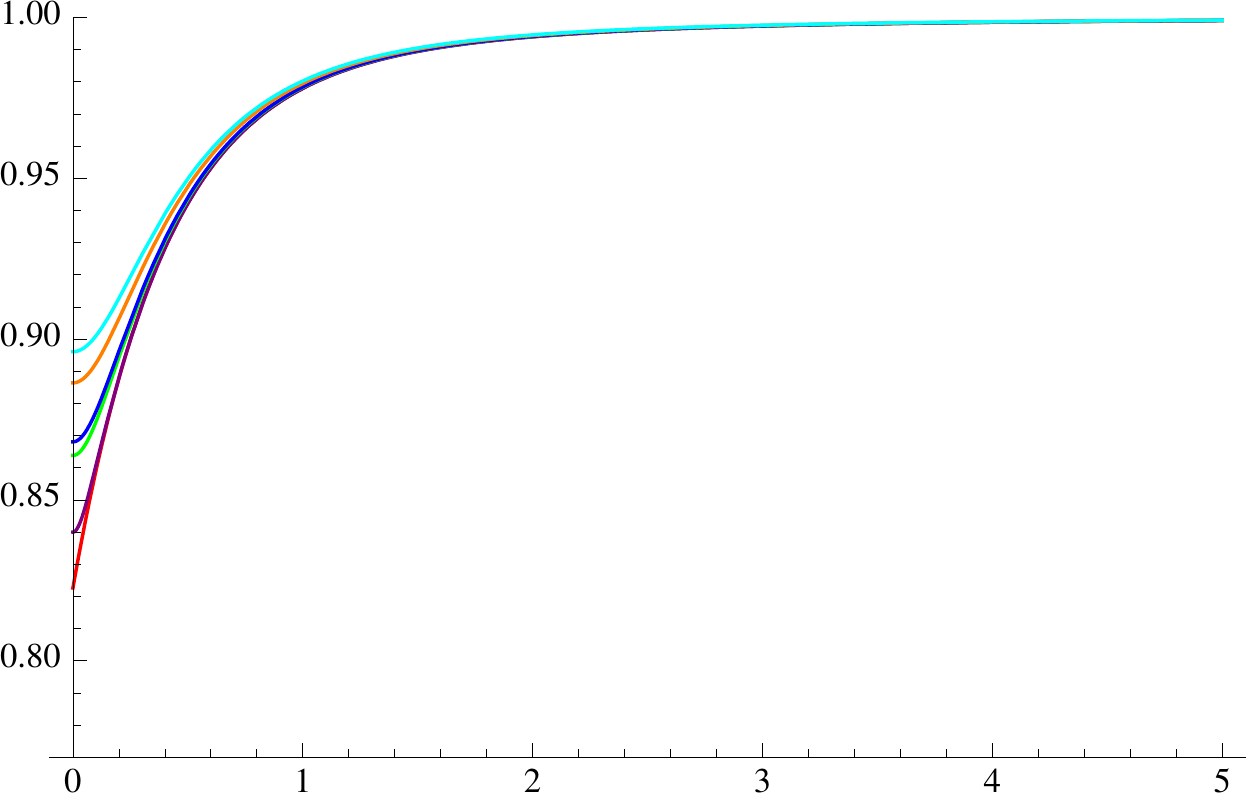}\hspace{5mm}
    \includegraphics[width=7cm, bb=0 0 360 232 ]{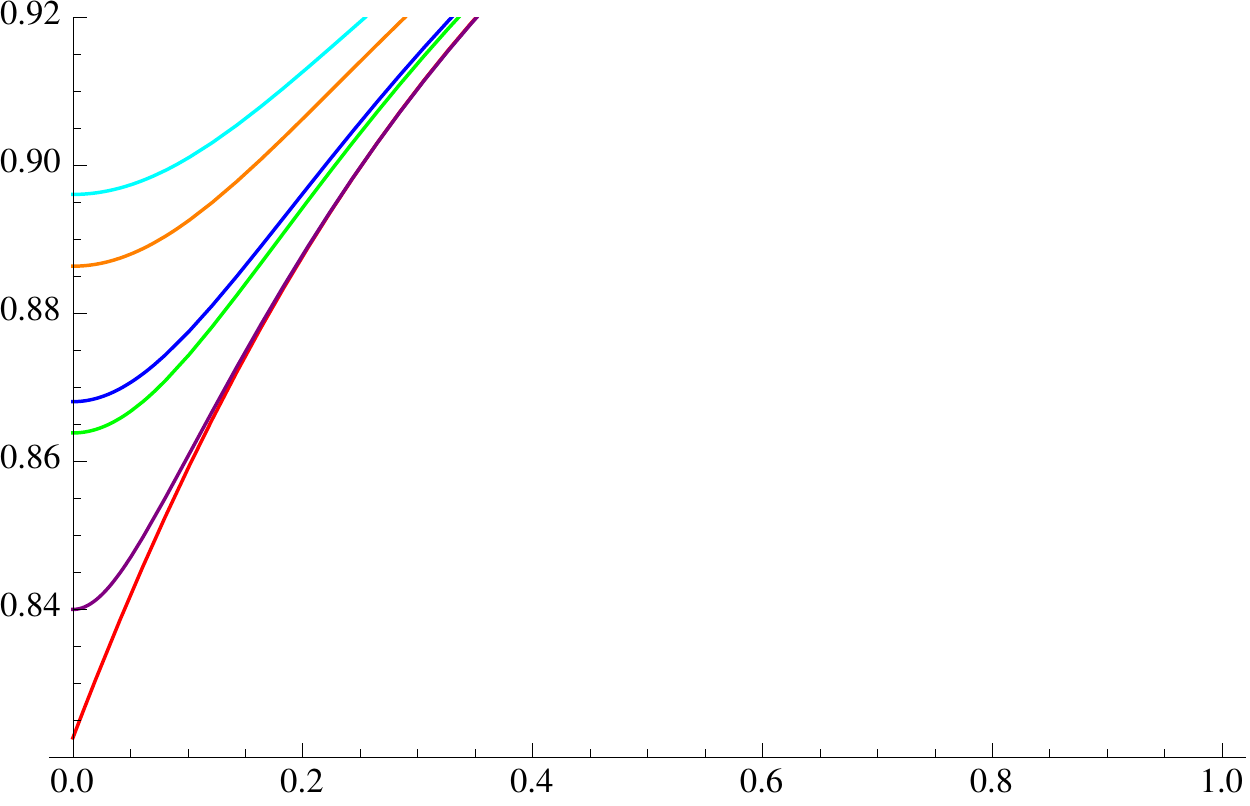}
    \put(-440,135){$w(\rho)$}
\put(-220,0){$\rho$}
\put(-210,135){$w(\rho)$}
\put(3,0){$\rho$}
\caption{The shape of the D7-brane with $|\vec{\mathcal{E}}|^2-|\vec{\mathcal{B}}|^2=0$. The curves correspond respectively to $\vec{\mathcal{E}}\cdot \vec{\mathcal{B}}/(\vec{\mathcal{E}}\cdot \vec{\mathcal{B}})_\textrm{cr}$ = 0.98, 0.99, 1, 1.001, 1.001, 1 from top to bottom and $(\vec{\mathcal{E}}\cdot \vec{\mathcal{B}})_\textrm{cr}\approx0.45793$. Right figure is an enlarged view around $\rho=0$.  }
\label{D3-D7fig1}
  \end{center}
  \end{figure}

\begin{figure}[thpb]
  \begin{center}
    \includegraphics[width=7cm,bb=0 0 360 232 ]{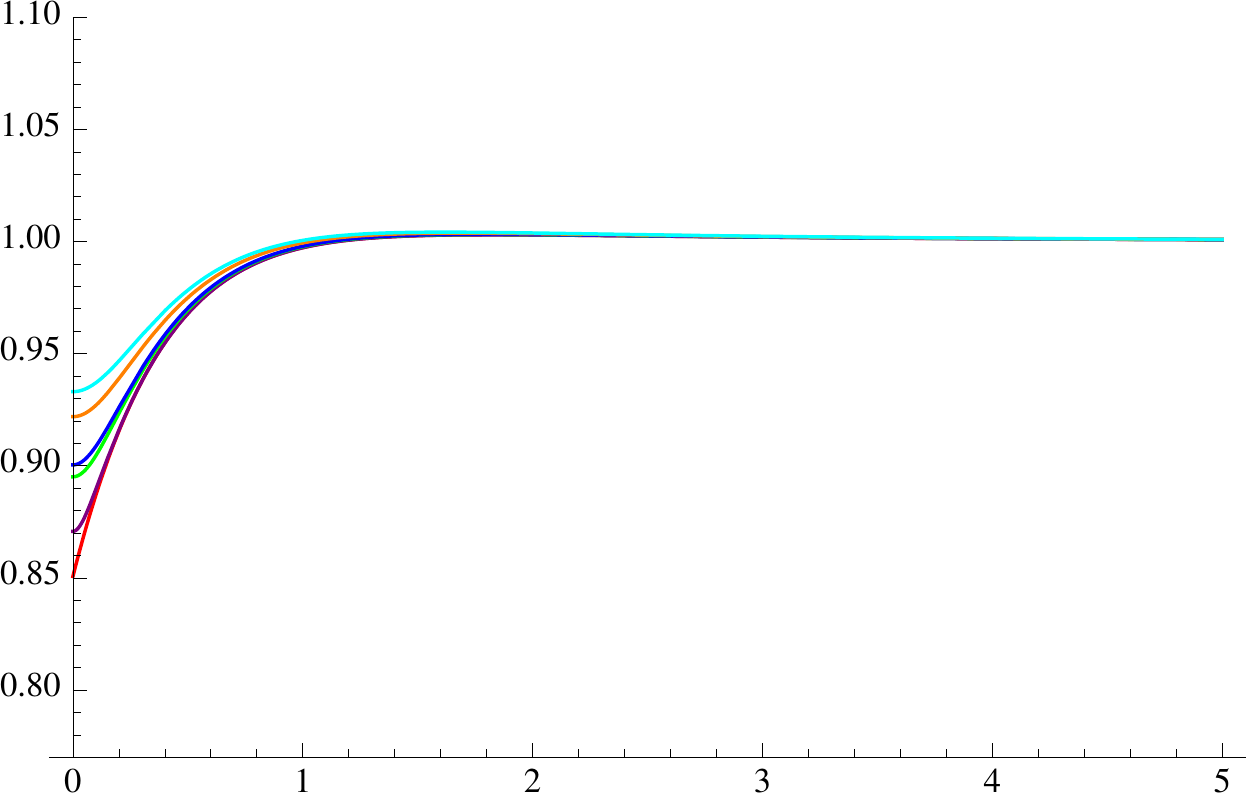}\hspace{5mm}
    \includegraphics[width=7cm, bb=0 0 360 232 ]{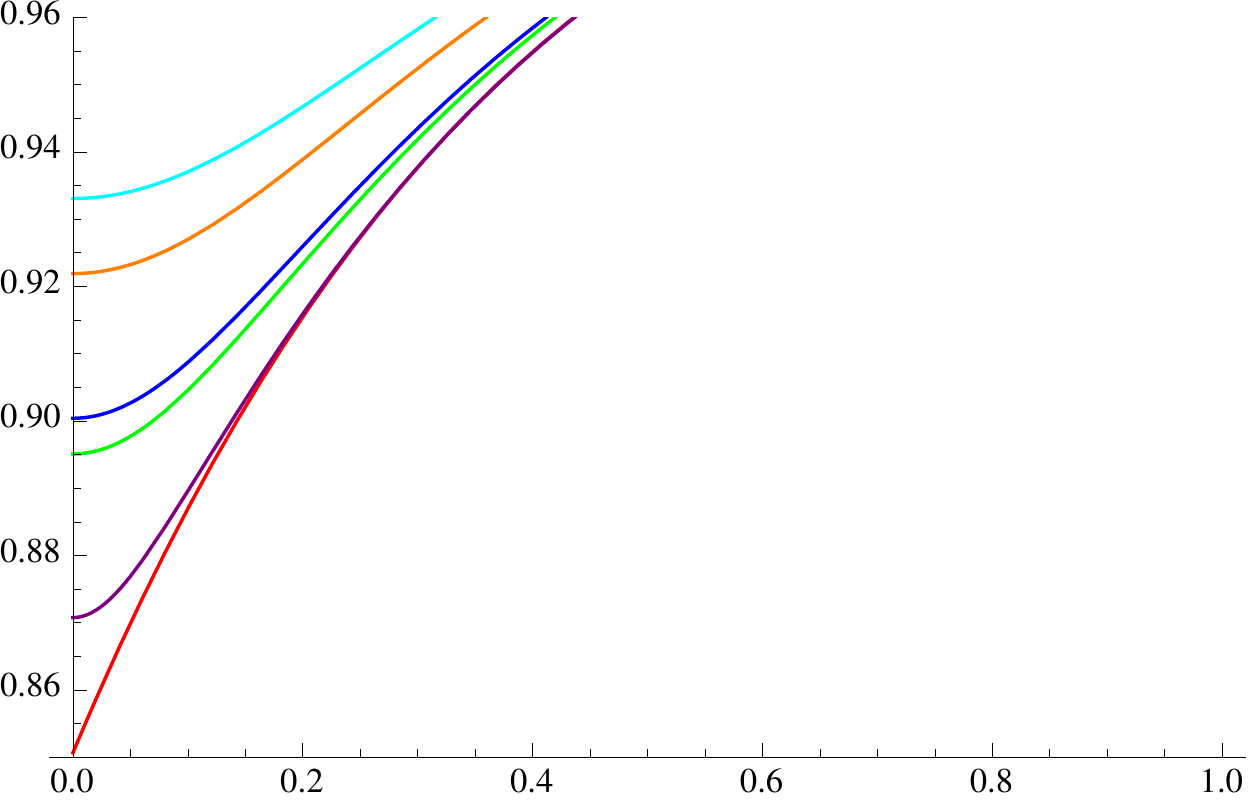}
    \put(-440,135){$w(\rho)$}
\put(-220,0){$\rho$}
\put(-210,135){$w(\rho)$}
\put(3,0){$\rho$}
\caption{The shape of the D7-brane with $|\vec{\mathcal{E}}|^2-|\vec{\mathcal{B}}|^2=-0.25$. The curves correspond respectively to $\vec{\mathcal{E}}\cdot \vec{\mathcal{B}}/(\vec{\mathcal{E}}\cdot \vec{\mathcal{B}})_\textrm{cr}$ = 0.98, 0.99, 1, 1.001, 1.001, 1 from top to bottom and $(\vec{\mathcal{E}}\cdot \vec{\mathcal{B}})_\textrm{cr}\approx0.63647$. Right figure is an enlarged view around $\rho=0$.  }
\label{D3-D7fig2}
  \end{center}
  \end{figure}

\begin{figure}[thpb]
  \begin{center}
    \includegraphics[width=7cm,bb=0 0 360 232 ]{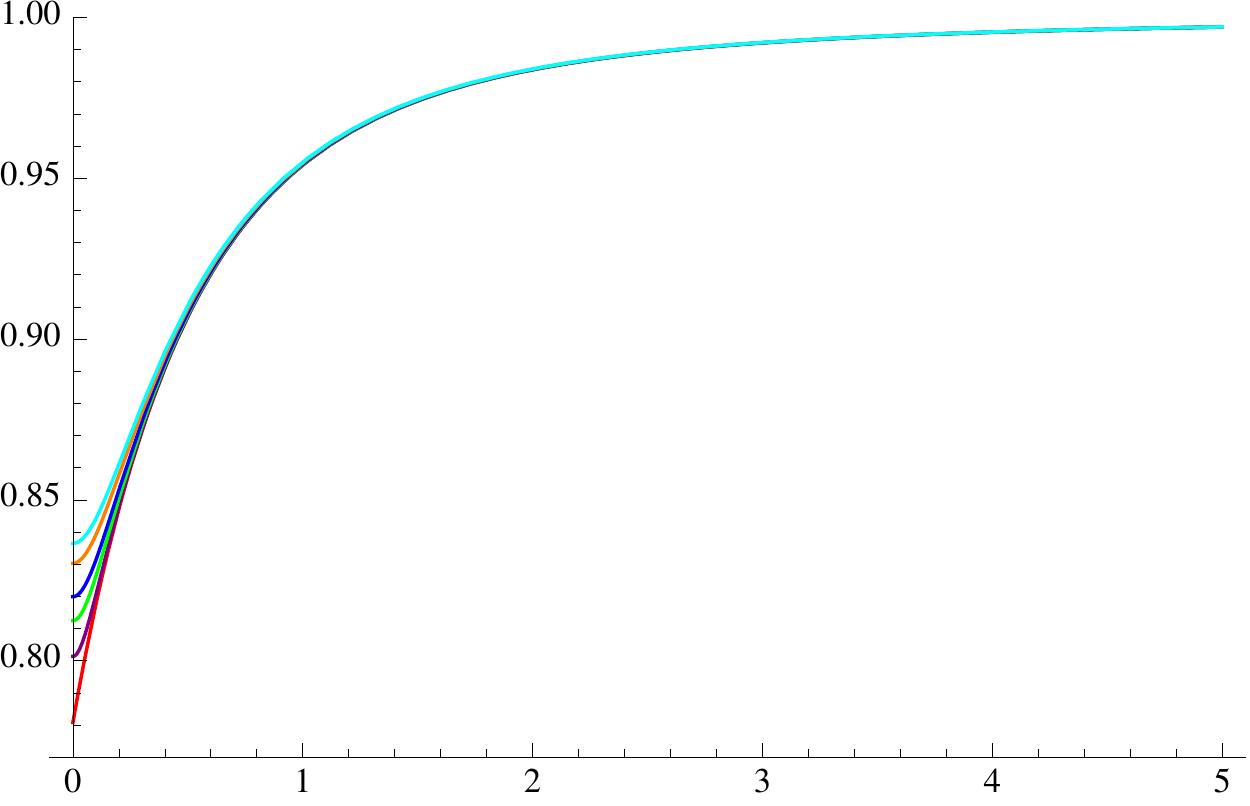}\hspace{5mm}
    \includegraphics[width=7cm, bb=0 0 360 232 ]{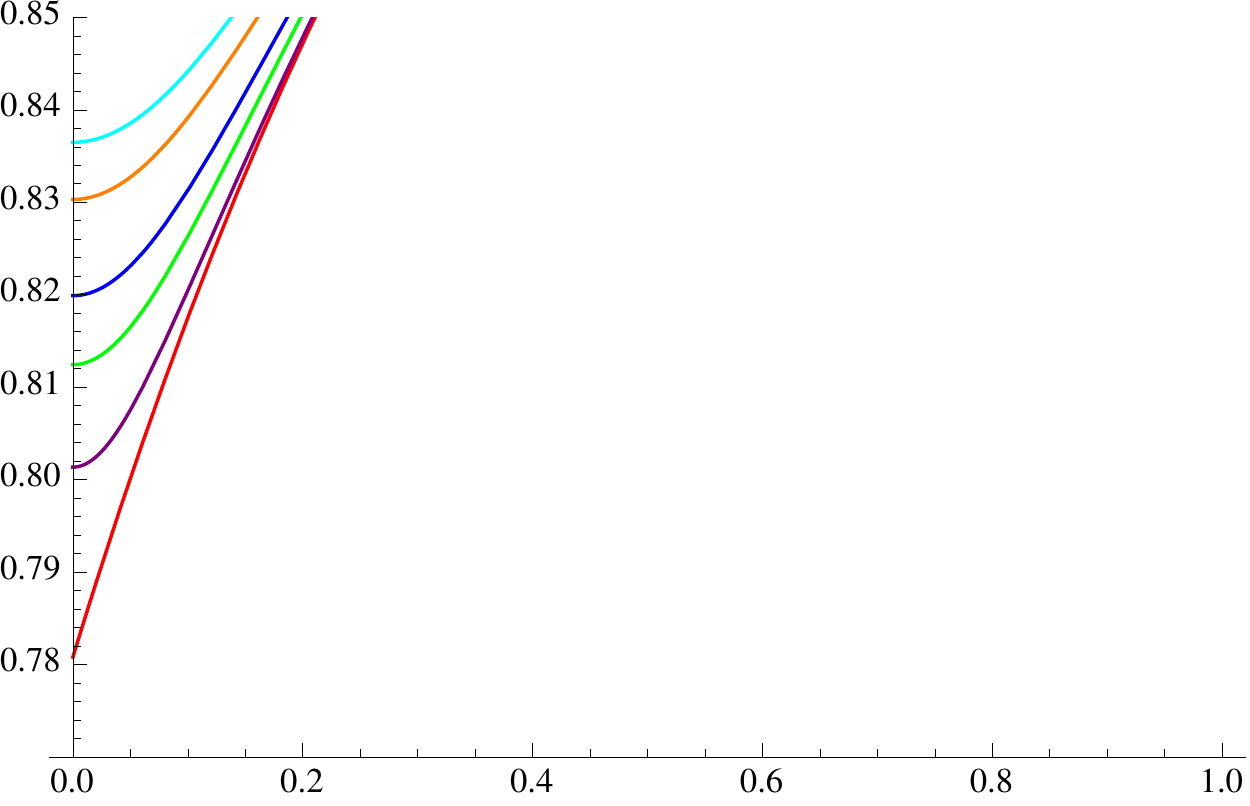}
    \put(-440,135){$w(\rho)$}
\put(-220,0){$\rho$}
\put(-210,135){$w(\rho)$}
\put(3,0){$\rho$}
\caption{The shape of the D7-brane with $|\vec{\mathcal{E}}|^2-|\vec{\mathcal{B}}|^2=+0.25$. The curves correspond respectively to $\vec{\mathcal{E}}\cdot \vec{\mathcal{B}}/(\vec{\mathcal{E}}\cdot \vec{\mathcal{B}})_\textrm{cr}$ = 0.98, 0.99, 1, 1.003, 1.003, 1 from top to bottom and $(\vec{\mathcal{E}}\cdot \vec{\mathcal{B}})_\textrm{cr}\approx0.21273$. Right figure is an enlarged view around $\rho=0$. }
\label{D3-D7fig3}
  \end{center}
  \end{figure}

In Fig.~\ref{D3-D7fig1}, Fig.~\ref{D3-D7fig2} and Fig.~\ref{D3-D7fig3},
we plot numerical solutions of $w(\rho)$, for given Lorentz-invariant combinations 
$|\vec{\mathcal{E}}|^2-|\vec{\mathcal{B}}|^2$ and $\vec{\mathcal{E}}\cdot \vec{\mathcal{B}}$. We find the critical embedding (conical D7-brane configurations)
which are of our interest.  In each figure, the red curves are the critical embeddings.

Fig.~\ref{D3-D7fig1}, Fig.~\ref{D3-D7fig2} and Fig.~\ref{D3-D7fig3} correspond
to different values of $|\vec{\mathcal{E}}|^2-|\vec{\mathcal{B}}|^2$ respectively.
In each figure, for those fixed values of $|\vec{\mathcal{E}}|^2-|\vec{\mathcal{B}}|^2$,
we vary $\vec{\mathcal{E}}\cdot \vec{\mathcal{B}}$ and find the critical embedding
(red curves in each figure). We define $(\vec{\mathcal{E}}\cdot \vec{\mathcal{B}})_\textrm{cr}$ as the value of $\vec{\mathcal{E}}\cdot \vec{\mathcal{B}}$ which gives 
the conical solutions.\footnote{
Note that there are solutions whose absolute value of $\vec{\mathcal{E}}\cdot \vec{\mathcal{B}}$ is larger than that of $(\vec{\mathcal{E}}\cdot \vec{\mathcal{B}})_\textrm{cr}$. Furthermore, for given electromagnetic fields, there may be several brane solutions. It is expected that those solutions are continuously connected
to the critical embedding in the parameter space, 
with a fractal-like structure (see some
discussions in Ref.~\cite{Erdmenger:2007bn}). Here we are just interested in the
critical embedding.
}

\begin{figure}[thpb]
  \begin{center}
    \includegraphics[width=7cm,bb=0 0 360 232 ]{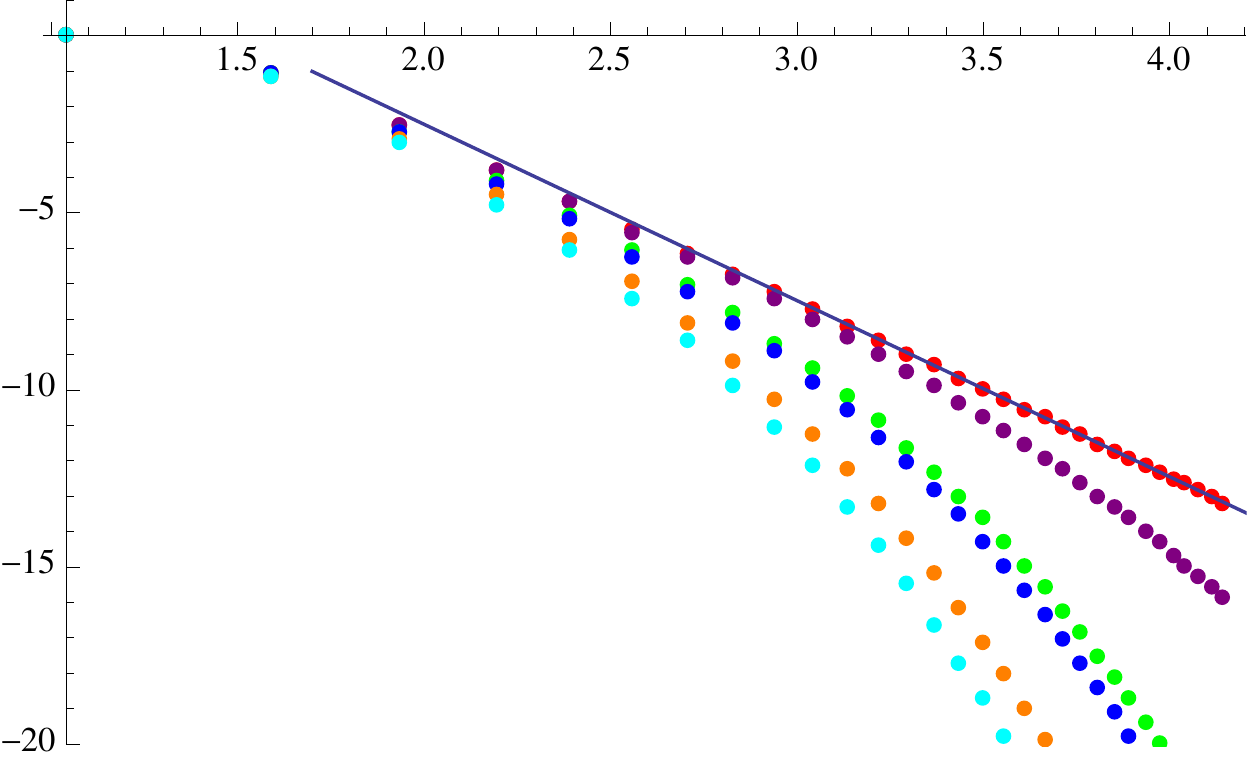}
 \put(-230,55){$\log{\frac{\varepsilon_n}{\varepsilon_0}}$}
\put(5,115){$\log{\frac{\omega_n}{m}}$ }
\put(-50,70){$\varepsilon_n\propto\omega_n^{-4.98}$ }
\caption{Plot of $\log{(\varepsilon_n/\varepsilon_0)}$ as a function of $\log{(\omega_n/m)}$ with $|\vec{\mathcal{E}}|^2-|\vec{\mathcal{B}}|^2=0$. The color of dotted lines corresponds with that of Fig.~\protect\ref{D3-D7fig1}'s curves. Black line is a fitting line of red dots at large $\omega_n$.}
\label{D3-D7fig4}
  \end{center}
  \end{figure}

\begin{figure}[thpb]
  \begin{center}
    \includegraphics[width=7cm,bb=0 0 360 232 ]{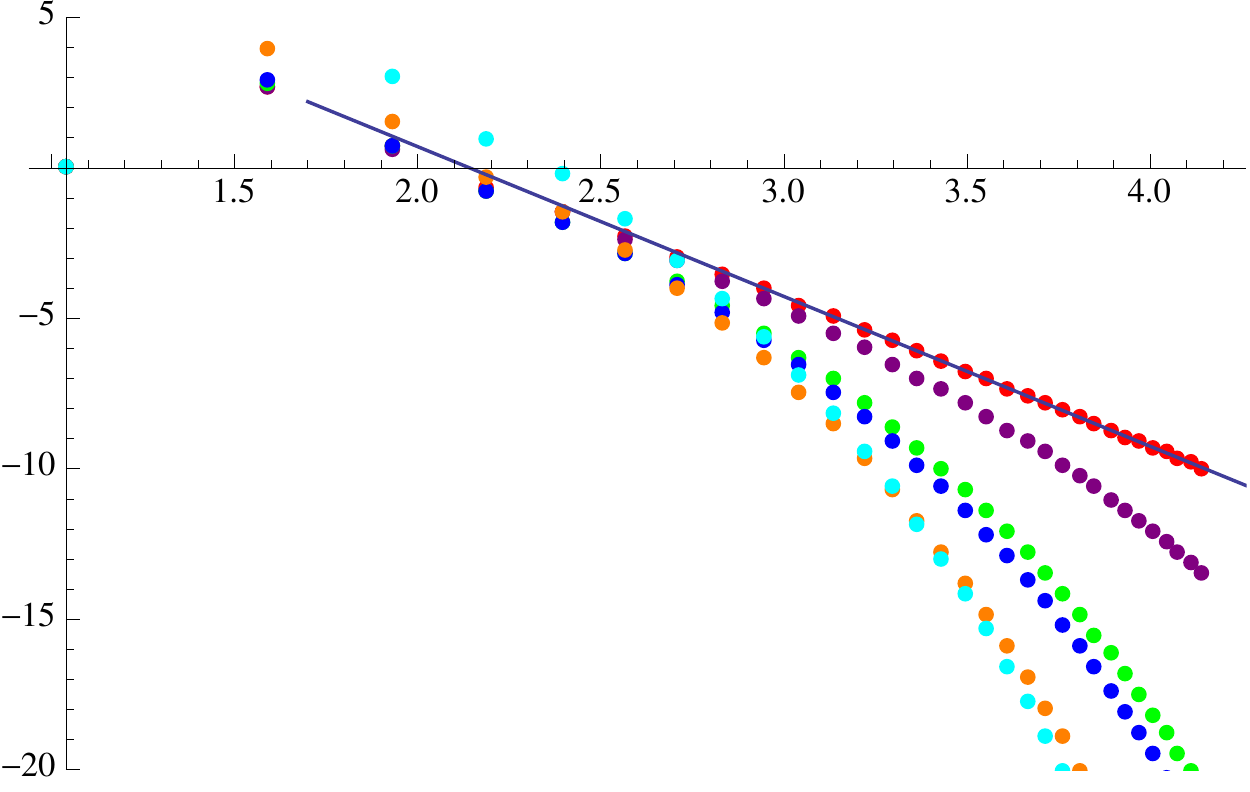}
 \put(-230,50){$\log{\frac{\varepsilon_n}{\varepsilon_0}}$}
\put(5,100){$\log{\frac{\omega_n}{m}}$ }
\put(-50,75){$\varepsilon_n\propto\omega_n^{-4.98}$ }
\caption{Plot of $\log{(\varepsilon_n/\varepsilon_0)}$ as a function of $\log{(\omega_n/m)}$ with $|\vec{\mathcal{E}}|^2-|\vec{\mathcal{B}}|^2=-0.25$. The color of dotted lines corresponds with that of Fig.~\protect\ref{D3-D7fig2}'s curves. Black line is a fitting line of red dots at large $\omega_n$.
}
\label{D3-D7fig5}
  \end{center}
  \end{figure}

\begin{figure}[thpb]
  \begin{center}
    \includegraphics[width=7cm,bb=0 0 360 232 ]{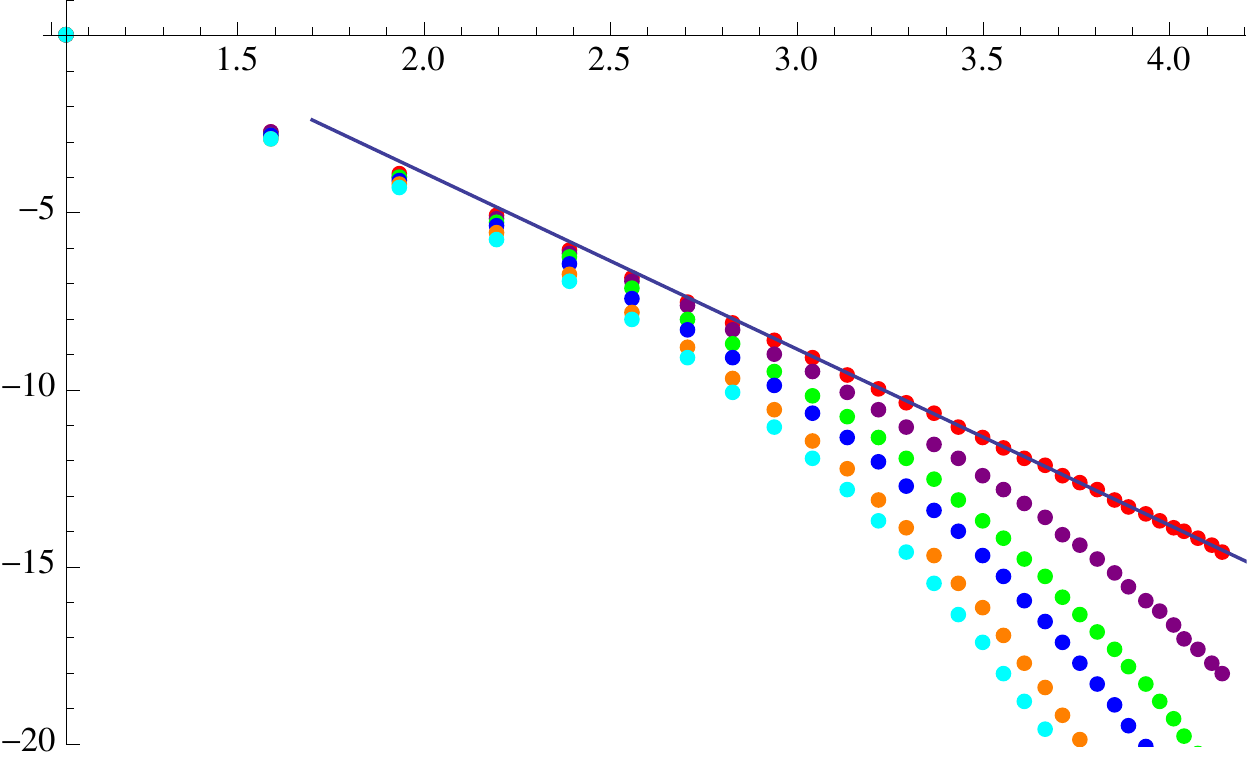}
 \put(-230,55){$\log{\frac{\varepsilon_n}{\varepsilon_0}}$}
\put(5,115){$\log{\frac{\omega_n}{m}}$ }
\put(-50,60){$\varepsilon_n\propto\omega_n^{-4.98}$ }
\caption{Plot of $\log{(\varepsilon_n/\varepsilon_0)}$ as a function of $\log{(\omega_n/m)}$ with $|\vec{\mathcal{E}}|^2-|\vec{\mathcal{B}}|^2=+0.25$. The color of dotted lines corresponds with that of Fig.~\protect\ref{D3-D7fig3}'s curves. Black line is a fitting line of red dots at large $\omega_n$.}
\label{D3-D7fig6}
  \end{center}
  \end{figure}

Finally we present the power-law of the energy distribution.
Fig.~\ref{D3-D7fig4}, Fig.~\ref{D3-D7fig5} and Fig.~\ref{D3-D7fig6} are plots of $\log{(\varepsilon_n/\varepsilon_0)}$ as a function of $\log{(\omega_n/m)}$. 
We notice that 
the red dotted lines (which are for the critical embeddings) are described well
by a linear fit, in particular at larger $\omega_n$ regions, in each figure,
while for other smooth embeddings the fit cannot be linear and the energy 
decrease rapidly for larger $\omega_n$.\footnote{
A peculiar feature of Fig.~\ref{D3-D7fig5} is that $\varepsilon_1$ is larger than 
$\varepsilon_0$. This is related with the shape of the D7-brane in 
Fig.~\ref{D3-D7fig2} where $\partial_\rho w(\rho)$ is negative at large $\rho$,
which shows the repulsion due to the worldvolume magnetic field.
}
The linear fit for the red dots 
means the power-law of the energy distribution which we expected.
Black lines are fitting lines of the red dots at large $\omega_n$. 
From the slope of the black line, 
we obtain the power-law scaling of the energy distribution as
\begin{align}
 \varepsilon_n\propto\omega_n^{-4.98}.
 \label{4.98}
 \end{align}
Since the power appearing in the fit does not depend on $|\vec{\mathcal{E}}|^2-|\vec{\mathcal{B}}|^2$ of our choices, the power is universal, and coincides with
$\alpha=-5$ given in Ref.~\cite{Hashimoto:2014dda, Hashimoto:2014xta}.


\subsection{Turbulence at a finite temperature}

In the previous subsection, we have found the universal power law (\ref{4.98}).
Here, instead, let us turn on a temperature rather than the electromagnetic field
to find the universality of the power law.
We introduce a nonzero temperature following \cite{Mateos:2007vn}, and calculate the power-law scaling of the energy distribution at the critical embedding. 

A nonzero temperature can be easily introduced by replacing the 
background AdS geometry by the AdS black hole metric as
\begin{align}
ds^2=\frac{r^2}{R^2}[-f(r)dt^2+d\vec{x}^2]+\frac{R^2}{r^2}\left[\frac{dr^2}{f(r)}+r^2d\Omega^2_5\right],\label{AdSblackholemetric}
\end{align}
where $f(r)\equiv1-(r_H/r)^4$ and the location of the horizon $r_H$ is related  to the temperature $T$ by the AdS/CFT dictionary as $T=r_H/\pi R^2$. We introduce a
new coordinate $u$ as
\begin{align}
2u^2=r^2+\sqrt{r^4-r^4_H},
\end{align}
with which the AdS black hole metric (\ref{AdSblackholemetric}) is written by
\begin{align}
ds^2&=\frac{u^2}{R^2}\left[-\frac{f(u)^2}{\tilde{f}(u)}dt^2+\tilde{f}(u)d\vec{x}^2\right]+\frac{R^2}{u^2}[du^2+u^2d\Omega_5^2]\notag\\
&=\frac{u^2}{R^2}\left[-\frac{f(u)^2}{\tilde{f}(u)}dt^2+\tilde{f}(u)d\vec{x}^2\right]+\frac{R^2}{u^2}[dv^2+v^2d\Omega_3^2+dw^2+d\bar{w}^2],\label{AdSblackholemetric2}
\end{align}
where 
\begin{align}
f(u)\equiv1-\frac{r_H^4}{4u^4},\;\;\tilde{f}(u)\equiv1+\frac{r_H^4}{4u^4},\;\;u^2=v^2+w^2+\bar{w}^2.
\end{align}

As in Sec.~\ref{sec:section2.2}, we use the ansatz
\begin{align}
w=w(v),\;\;\bar{w}=A_a=0,
\end{align}
and a boundary condition 
\begin{align}
w(v=\infty)=R^2m.\label{D3-D7bc2}
\end{align} 
Substituting our ansatz (\ref{AdSblackholemetric2}) to (\ref{DBI D3-D7}), we obtain the flavor D7-brane action as
\begin{align}
S=\int dtd^3x\int^\infty_0dvv^3\left(1-\frac{r_H^8}{16u^8}\right)\sqrt{1+(\partial_vw)^2}.
\end{align}
The Euler-Lagrange equation for $w(v)$ is 
\begin{align}
0=\partial_v\left[v^3\left(1-\frac{r_H^8}{16(v^2+w^2)^4}\right)\frac{\partial_vw}{\sqrt{1+(\partial_vw)^2}}\right]-\frac{v^3wr_H^8}{2(v^2+w^2)^5}\sqrt{1+(\partial_vw)^2}.\label{D3-D&eqwithtemperature}
\end{align}
Following the strategy of \cite{Hashimoto:2014xta,Hashimoto:2014dda} and that in Sec.~\ref{sec:section2.2}, we calculate $w(v)$ and $\varepsilon_n$ from (\ref{D3-D&eqwithtemperature}) with (\ref{D3-D7bc2}) numerically. For the numerical simulations we set $R=m=1$ as before.

\begin{figure}[thpb]
  \begin{center}
    \includegraphics[width=7cm,bb=0 0 360 232 ]{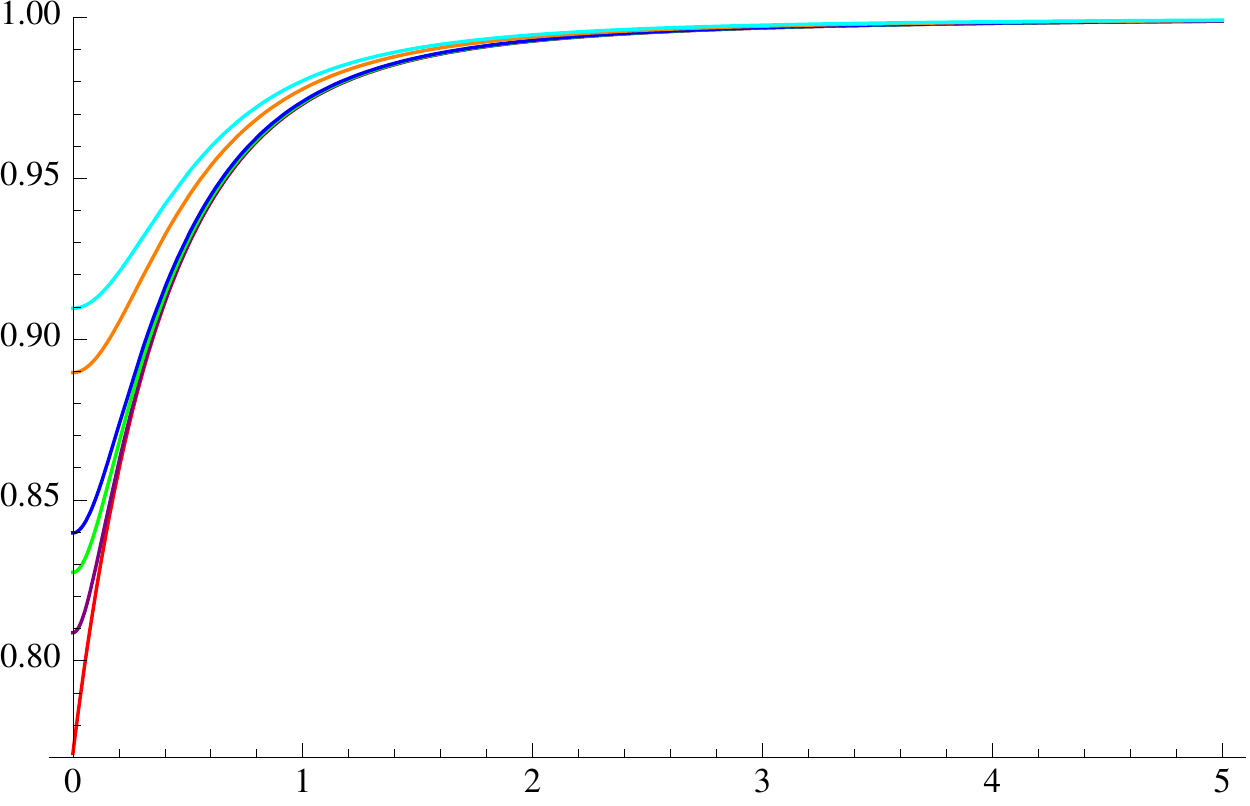}\hspace{5mm}
    \includegraphics[width=7cm, bb=0 0 360 232 ]{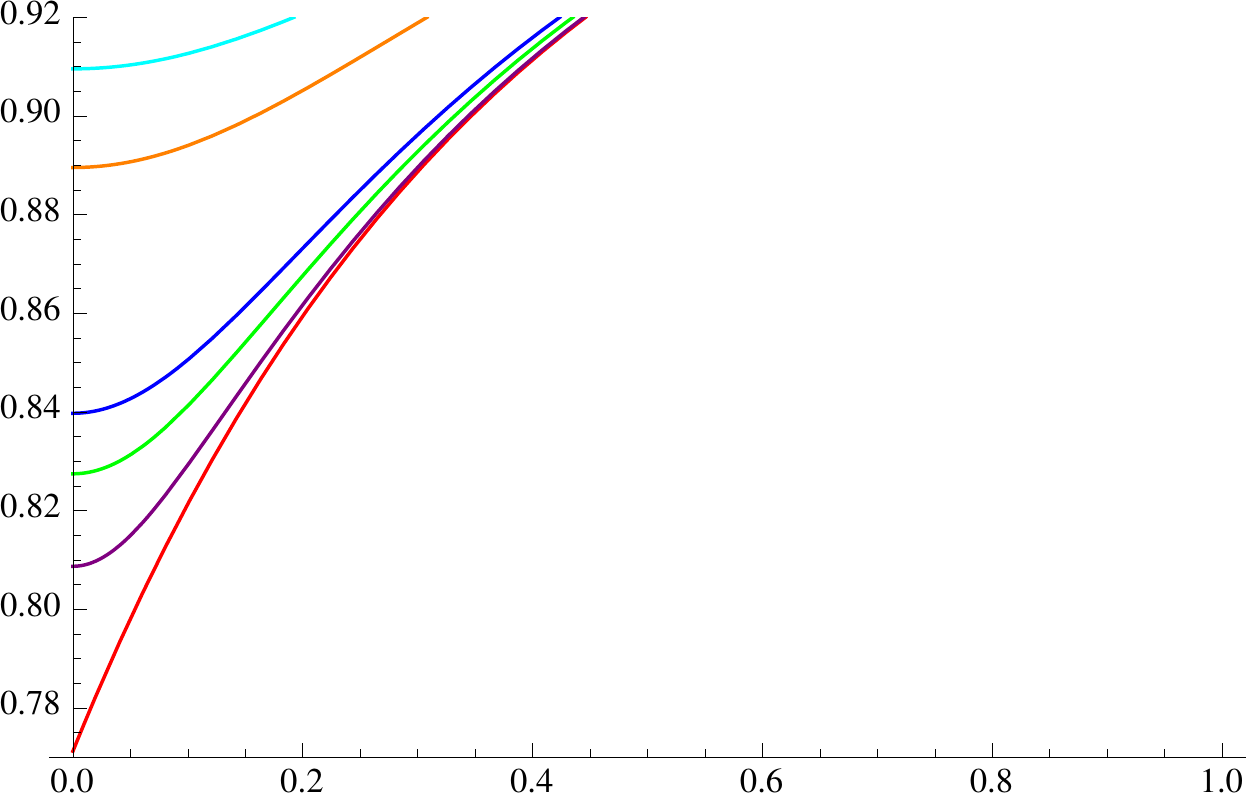}
  \put(-440,135){$w(v)$}
\put(-215,0){$v$}
\put(-210,135){$w(v)$}
\put(3,0){$v$}
\caption{The shape of the D7-brane with the finite temperature effect. The curves correspond respectively to $r_H/(r_H)_\textrm{cr}$ = 0.98, 0.99, 1, 1.0005, 1.0005, 1 from top to bottom and $(r_H)_\textrm{cr}\approx1.0907$. Right figure is an enlarged view around $v=0$. }
\label{D3-D7fig7}
  \end{center}
  \end{figure}

Fig.~\ref{D3-D7fig7} describes the shape of the D7-brane with the finite temperature effect. To find a critical embedding which is a conical shape of the flavor D7-branes,
we vary the value of $r_H$ while fixing the quark mass $m=1$. The red conical solution in Fig~\ref{D3-D7fig7} is the one which we want. 
We define $(r_H)_\textrm{cr}$ as the value of $r_H$ which provides the conical solution. In the figure we also plot other solutions of Minkowski embedding for comparison.\footnote{Note that lowering the tip of the D7-brane does not directly corresponds to raising the temperature. For a certain region of values of $r_H$, 
there exists several solutions for a given value of $r_H$.  Furthermore, there are solutions whose $r_H$ is larger than $(r_H)_\textrm{cr}$. The situation is similar to
that in the previous section.  The existence of such solutions was 
discussed in Ref.~\cite{Mateos:2007vn}.}

\begin{figure}[thpb]
  \begin{center}
    \includegraphics[width=7cm,bb=0 0 360 232 ]{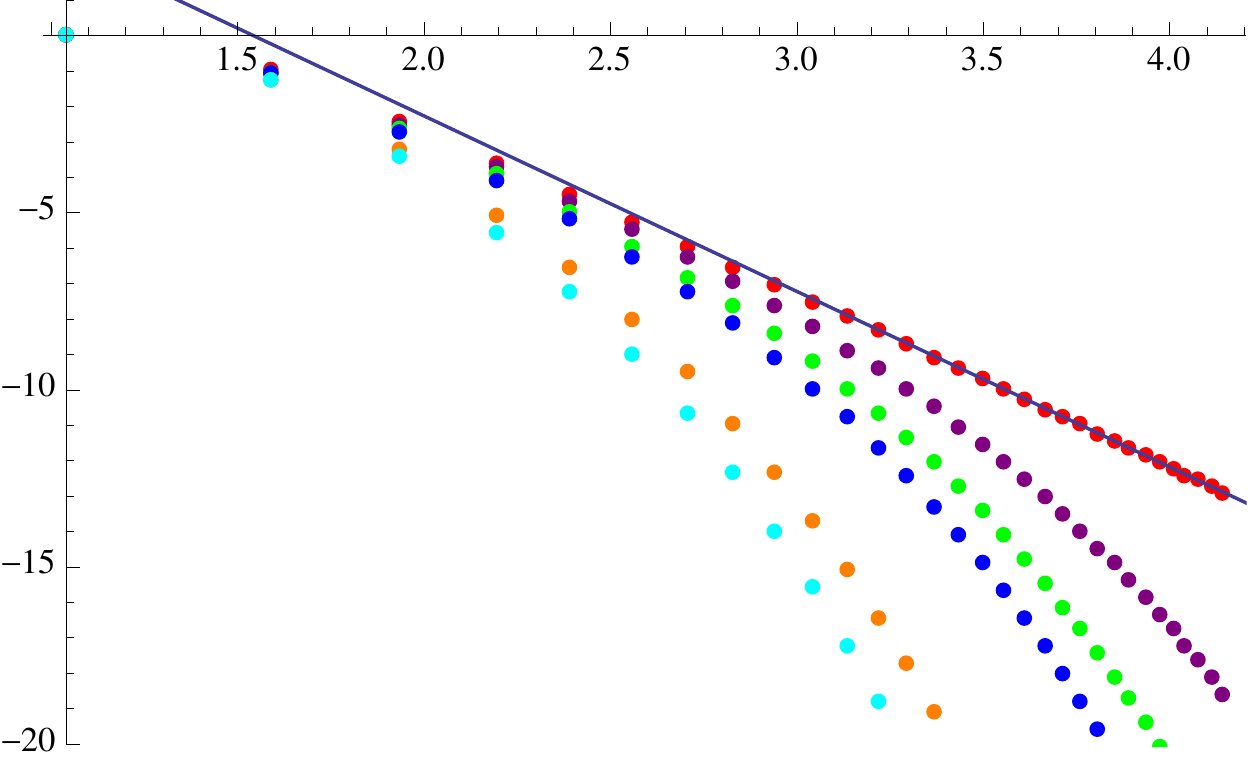}
 \put(-230,55){$\log{\frac{\varepsilon_n}{\varepsilon_0}}$}
\put(5,115){$\log{\frac{\omega_n}{m}}$ }
\put(-50,70){$\varepsilon_n\propto\omega_n^{-4.96}$ }
\caption{Plot of $\log{(\varepsilon_n/\varepsilon_0)}$ as a function of $\log{(\omega_n/m)}$ with the finite temperature effect. The colors of dotted lines correspond to those of Fig.~\protect\ref{D3-D7fig7}'s curves, respectively. The black line is a linear fit of the 
red dots at large $\omega_n$.}
\label{D3-D7fig8}
  \end{center}
  \end{figure}

Fig.~\ref{D3-D7fig8} is a plot of $\log{(\varepsilon_n/\varepsilon_0)}$ as a function of $\log{(\omega_n/m)}$, for our probe D7-brane at finite temperature.
The notation of the colors follow that of the previous section: 
the red dotted line (corresponding to the critical embedding) is found to obey
a linear fit at large $\omega_n$ region. It means a  power-law of the energy distribution for highly excited meson states at the critical embedding. 
The black line is a linear fit of the red dotted points at large $\omega_n$, 
and we find a power-law scaling again, 
\begin{align}
 \varepsilon_n\propto\omega_n^{-4.96}.
 \label{4.96}
 \end{align}
 
Summarizing the above, we have found the 
power-laws of the energy distribution for highly excited scalar meson states as (\ref{4.98}) and (\ref{4.96}).
The powers are very close to $\alpha = -5$. Thus we find the universality,\footnote{
We expect that the numerical power-law scaling factor will approach $-5$ if we 
extend our analysis to much higher states.}
\begin{align}
\varepsilon_n\propto\omega_n^{-5} \, .
\end{align}
The power scaling is universal at critical embedding, and does not depend on
how the critical embedding is realized: 
either by various electromagnetic fields or a finite temperature.

\section{D3-D5 system and universal turbulence}

In this section, we analyze the turbulent behavior of the mesons at excited states in D3-D5 brane system by using the AdS/CFT correspondence. 
The purpose is to find (1) whether we find the power-law behavior, and (2) 
whether the power is universal.

We study two ways to obtain the critical embedding of the probe D5-brane, as 
in the previous section. 
After reviewing the energy of the meson at  zero temperature and with 
no electromagnetic field, 
we introduce a constant electric field, and study how the relationship between 
the energy spectrum $\varepsilon_n$ and the mass $\Omega_n$ of the mesons is by changing the background values and accordingly the shape of the D5-brane.
In addition, we are interested also in the turbulent power law for the meson states at a finite temperature, and we examine the universality.

In this section we will find the energy distribution as $\varepsilon_n\propto(\Omega_n)^{-4}$ at critical embeddings of the D5-brane, either for a constant electric field or a finite temperature system. The power is shared by both the situations, so is for
the D3-D5 system. However, 
the power itself is $-4$ which is different from the value of the power, $-5$,
for the D3-D7 system.

\subsection{Review of the $\mathcal{N}=2$ supersymmertic defect gauge 
theory in AdS/CFT}
In this subsection, we review the derivation \cite{Myers:2006qr}
of the spectrum of ``mesons"  
from the fluctuation of a probe D5-brane at zero temperature and with
no background electric field, by using the AdS/CFT correspondence.
We are interested in the following D3-D5 brane configuration:
\\

\hspace{20mm}
\begin{tabular}{|c|c|c|c|c|c|c|c|c|c|c|} 
\hline
      & 0 & 1 & 2 & 3 & 4 & 5 & 6 & 7 & 8 & 9 \\ \hline
D3 & $\surd$ & $\surd$ &  $\surd$ & $\surd$ &  &  &  &  &  &   \\ \hline
D5 & $\surd$ & $\surd$ & $\surd$ &  & $\surd$ & $\surd$ & $\surd$ &\hspace{3mm} &\hspace{3mm} & \hspace{3mm}  \\ 
\hline 
\end{tabular}
\\ 

\noindent
The brane configuration preserves $\mathcal{N}=2$ supersymmetries in total. 
The flavor probe D5-brane extends along the directions $x^{0},x^{1},x^{2},x^{4},x^{5},x^{6}$, so it shares only $x^0,x^1,x^2$ directions with 
the gauge $N_c$ D3-branes.  
Thus, 
the gluon ${\cal N}=4$ multiplets live in the (3+1) dimensional spacetime while the quark hypermultiplets (and resultantly the mesons as their bound states) 
live only in a (2+1) dimensional domain wall, which is a defect.

It is known that meson states can be analyzed holographically by
the D3-D5 brane system. 
The scalar meson field corresponds to a fluctuation of the probe D5-brane along the directions  $x^{7},x^{8},x^{9}$, due to the AdS/CFT correspondence.
For the fluctuation, the Laplace equation is classically solved by a Gauss hypergeometoric function \cite{Myers:2006qr}, as we will see below.

The $\mathrm{AdS}_5\times \mathrm{S}^5$ background metric is
\begin{align}
ds^{2} = \frac{r^2}{R^2}\eta_{\mu\nu}dx^\mu dx^\nu + \frac{R^2}{r^2}[d\rho^2 + \rho^2d\Omega^{2}_{2} + d\omega^2_4 + d\omega^2_5 + d\omega^2_6].\label{AdS}
\end{align}
The 5-sphere together with the AdS radial direction $r$ is decided into
a combination of $\rho$ and 2-sphere $\Omega_2$ and the remaining 
three directions 
$x^7, x^8, x^9$ which are parametrized by $\omega_4, \omega_5, \omega_6$. 
So the relation among these parameters is 
$r^2=\rho^2+\omega_4^2+\omega_5^2+\omega_6^2$.
Since $(\omega_4, \omega_5, \omega_6)$ has a rotation symmetry, we 
may fix $\omega_5 = \omega_6 = 0$ for simplicity. 

The D5-brane action is a Dirac-Born-Infeld(DBI) action given by
\begin{align}
S_{\mathrm{DBI}} = - \tau_5\int{d^6\xi}\ e^{-\phi}\sqrt{-{\rm det}(P[g]_{ab} + 2\pi l_{s}^2F_{ab})},\label{DBI}
\end{align}
where $\tau_5$ is the D5-brane tension and is defined by $\tau_5 = 1/(2\pi)^5g_s l_s^{6}$. $g_s$ is the string coupling. $\phi$ is the dilation field which is set to 
zero as the background. 

The scalar mesons are eigen modes which are liberalized solutions
for fluctuations of $\omega_4(x^i, \rho)$ of the D5-brane DBI action to which the 
background metric is substituted. Denoting the index $i$ running $i=0, 1, 2$
as the D5-brane does not extend along the direction $x^3$, we 
impose a boundary condition at the AdS boundary as
\begin{align}
\omega_4(x^i,\rho = \infty) = R^2m,
\end{align}
where $m$ is related to the quark mass $m_q$ as 
$m_q=(\lambda/2\pi^2)^{1/2}m$ due to
the AdS/CFT dictionary. 
A static classical solution of the DBI equation of motion is
\begin{align}
\omega_4(x^i,\rho) = R^2m.
\end{align}
The solution roughly measures the distance between the D3-branes and the D5-brane.

Next, we consider the fluctuation $\chi$ around the static solution $R^2m$,
defined by $\chi\equiv R^{-2}\omega_4 - m$. We assume for simplicity that
$\chi$ is independent of the coordinates $x^1$ and $x^2$.
The action for $\chi$ obtained by just expanding (\ref{DBI}) to the quadratic order
is
\begin{align}
S = \int{d^3x}\int^{\infty}_{0}{d\rho}\frac{\rho^2R^2m}{2(\rho^2+R^4m^2)^2}\left[(\partial_t\chi)^2 - \frac{(\rho^2+R^4m^2)^2}{R^4}(\partial_\rho\chi)^2\right] + \mathcal{O}(\chi^3),\label{S}
\end{align}
where the irrelevant overall factor is neglected. From (\ref{S}), we derive  the equation of motion as
\begin{align}
\left[\frac{\partial^2}{\partial t^2} - \frac{(\rho^2+R^4m^2)^2}{\rho^2R^2m}\frac{\partial}{\partial \rho}\frac{\rho^2m}{R^2}\frac{\partial}{\partial \rho} \right]\chi=0 ,
\end{align}
Its normalizable solution is
\begin{align}
\chi &= \sum^{\infty}_{n=0}{\rm Re}\left[C_n\exp[i\Omega_nt]E_{n}(\rho)\right],  
\end{align}
The basis functions are given by
\begin{align}
E_{n}(\rho)&\equiv \frac{4(n+1)}{\sqrt{\pi}}\left(\frac{R^4m^2}{\rho^2+R^4m^2}\right)^{n+\frac{1}{2}}F\left(-n,-1/2-n,3/2;-\frac{\rho^2}{R^4m^2}\right),\label{basis}
\end{align}
where $F$ is the Gaussian hypergeometric function. 
The mass of the level $n$ resonance meson is 
\begin{align}
\Omega_n &\equiv 2\sqrt{\left(1/2 +n\right)\left(3/2+n\right)} \, m .
\end{align}
For our later purpose we define
the inner product in the $\rho$-space as
\begin{align}
(F,G)\equiv \int^{\infty}_{0}{d\rho}
\frac{\rho^2R^2m}{(\rho^2+R^4m^2)^2}F(\rho)G(\rho), \label{inner}
\end{align}
under which we have the orthonormality condition
\begin{align}
(E_n,E_m)=\delta_{nm}.
\end{align}
Using the basis (\ref{basis}), we expand the fluctuation of the scalar field as
\begin{align}
\chi=\sum^{\infty}_{n=0}c_n(t)E_n(\rho),
\end{align}
then with this we can define the linearlized 
meson energy at level $n$, 
\begin{align}
\varepsilon_n\equiv \frac{1}{2}(\dot{c}_n^2+\Omega_n^2c_n^2). \label{D3-D5en}
\end{align}
The linearlized total energy is given by the formula (\ref{D3-D7totalen}). All of these are 
analogous to what we have used in the case of the D3-D7 brane system in Sec.~2.

\subsection{Turbulence with an electric field}

We are ready for studying the turbulent meson behavior. 
The turbulence should show up in the meson energy distribution 
which is defined as (\ref{D3-D5en}).
Note that the energy spectrum (\ref{D3-D5en}) is defined with no background electric field.
Once the electric field is turned on, the meson spectrum changes compared to that with no electric field, since the probe D5-brane is affected by the electric field.
It is well-known that the shape of the probe D-brane is deformed by the electric field \cite{Erdmenger:2007bn}.
We are interested in how the shape of the D5-brane changes and 
how the meson spectrum is affected accordingly.

\begin{figure}[htbp]
  \begin{center}
    \includegraphics[width=8.0cm,bb=0 0 842 595]{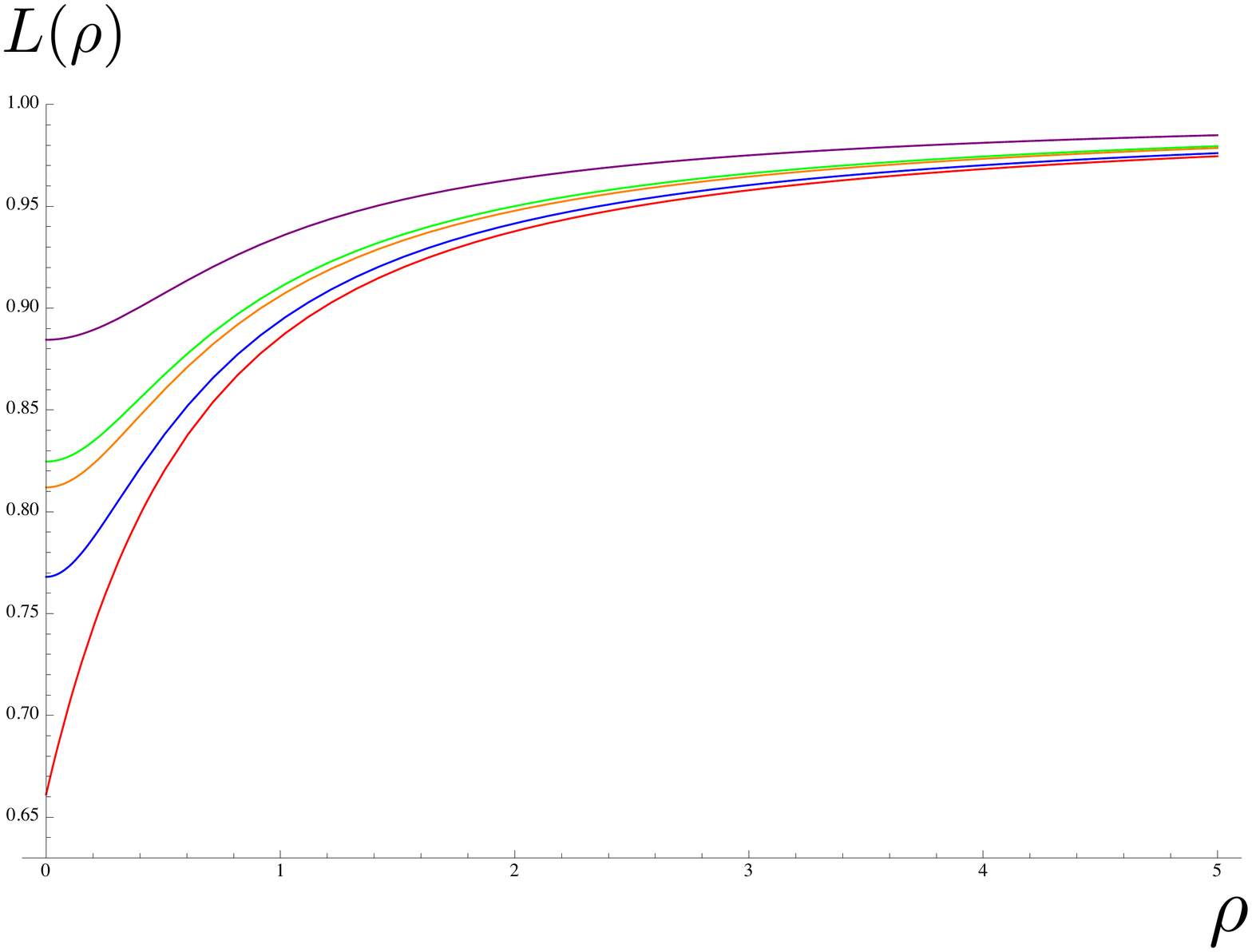}
    \caption{The shape of the D5-brane in the AdS background. 
Each curve corresponds, from top to bottom, to $\mathcal{E}/\mathcal{E}_{\mathrm{cr}}=0.9, 0.99,1,1.017,1$ respectively.
The D5-brane can have a cusp at $\rho=0$
for $\mathcal{E}=\mathcal{E}_{\mathrm{cr}}$$(\simeq 0.437)$
(the red curve). 
}
    \label{fig:D5shapeE}
  \end{center}
\end{figure}

To look at how the D5-brane shape is deformed, we solve the equation of motion
from the DBI action (\ref{DBI}).
$\mathrm{AdS}_5\times \mathrm{S}^5$ background is given by (\ref{AdS}).
From a simple consideration of rotational symmetry, we can put
$\omega_5,=\omega_6=0$
while $\omega_4$ depends on the radial direction $\rho$ as $w_4=L(\rho)$. 
Then, the induced metric on the D5-brane is given by
\begin{align}
ds^{2} = \frac{r^2}{R^2}(-dt^2 + \delta_{ij}dx^idx^j) + \frac{R^2}{r^2}[\{1+(\partial_\rho L)^2\}d\rho^2 + \rho^2d\Omega^{2}_{2}] \label{D5},
\end{align}
where $i,j=1,2$.
Then we turn on a constant electromagnetic field. In (2+1) dimensions, Lorentz 
transformation can bring us to a frame on which only $F_{01}$ component is
nonzero.
Denoting $E\equiv F_{01}$, 
we obtain the action with the electric field as\footnote{
The Chern-Simons terms in the D-brane effective actions does not contribute
in our analysis. For D4-D8 setup, they may contribute. See a discussion in
\cite{Hashimoto:2014yya}.}
\begin{align}
S_{\mathrm{DBI}} = -4\pi\tau_{5}\int{d^{3}x}\int^{\infty}_{0}{d\rho}\ \rho^2\sqrt{1+(\partial_{\rho}L)^2}\sqrt{1- \frac{R^4 (2\pi l_s^2)^2E^2}{(\rho^2 + L^2)^2}},
\end{align}
The equation of motion is obtained with a redefinition $\mathcal{E}\equiv 2\pi l_s^2E$
as
\begin{align}
\partial_\rho\left[\frac{\rho^2\partial_\rho L\sqrt{1- \frac{R^4\mathcal{E}^2}{(\rho^2 + L^2)^2}}}{\sqrt{1+(\partial_{\rho}L)^2}} \right] - \frac{2R^4 \mathcal{E}^2\rho^2L\sqrt{1+(\partial_{\rho}L)^2}}{(\rho^2+L^2)^3\sqrt{1- \frac{R^4\mathcal{E}^2}{(\rho^2 + L^2)^2}}} = 0.
\end{align}
We solve this equation numerically to obtain the shape of the probe D5-brane
for a given value of $\mathcal{E}$.
The results of the numerical calculations, with $R=1$ and $m=1$, are shown in Fig.~\ref{fig:D5shapeE}.
the shape of the D5-brane changes and the D5-brane has a cusp as the static electric field increases. With various chosen background electric field $\mathcal{E}$, the D5-brane
changes its shape, and we find a conical D5-brane. It is the critical embedding, 
for which we expect the turbulent behavior of mesons.

Basically, as in the case of the D7-brane probe, for small $\mathcal{E}$ the
D5-brane is at the Minkowski embedding (which has a smooth shape), while for a large enough $\mathcal{E}$ the D5-brane is at a black hole embedding where the induced 
metric on the D5-brane has an effective horizon. The critical embedding is in between
those embeddings, and the D5-brane has a cone at the tip.
The value of $\mathcal{E}$ with which the D5-brane can be in the critical embedding 
is called $\mathcal{E}_{\mathrm{cr}}$, and in Fig.~\ref{fig:D5shapeE} we plot various
curves for various $\mathcal{E}/\mathcal{E}_{\mathrm{cr}}$.\footnote{Note again that the
location of the tip of the D5-brane is not a monotonic function of $\mathcal{E}/\mathcal{E}_{\mathrm{cr}}$. In some region of the value of $\mathcal{E}/\mathcal{E}_{\mathrm{cr}}$,
the D5-brane profile is found to be not unique. A fractal-like structure emerges, as
in the case of the D3-D7 system.}

\begin{figure}[htbp]
  \begin{center}
    \includegraphics[width=8.0cm,bb=0 0 842 595]{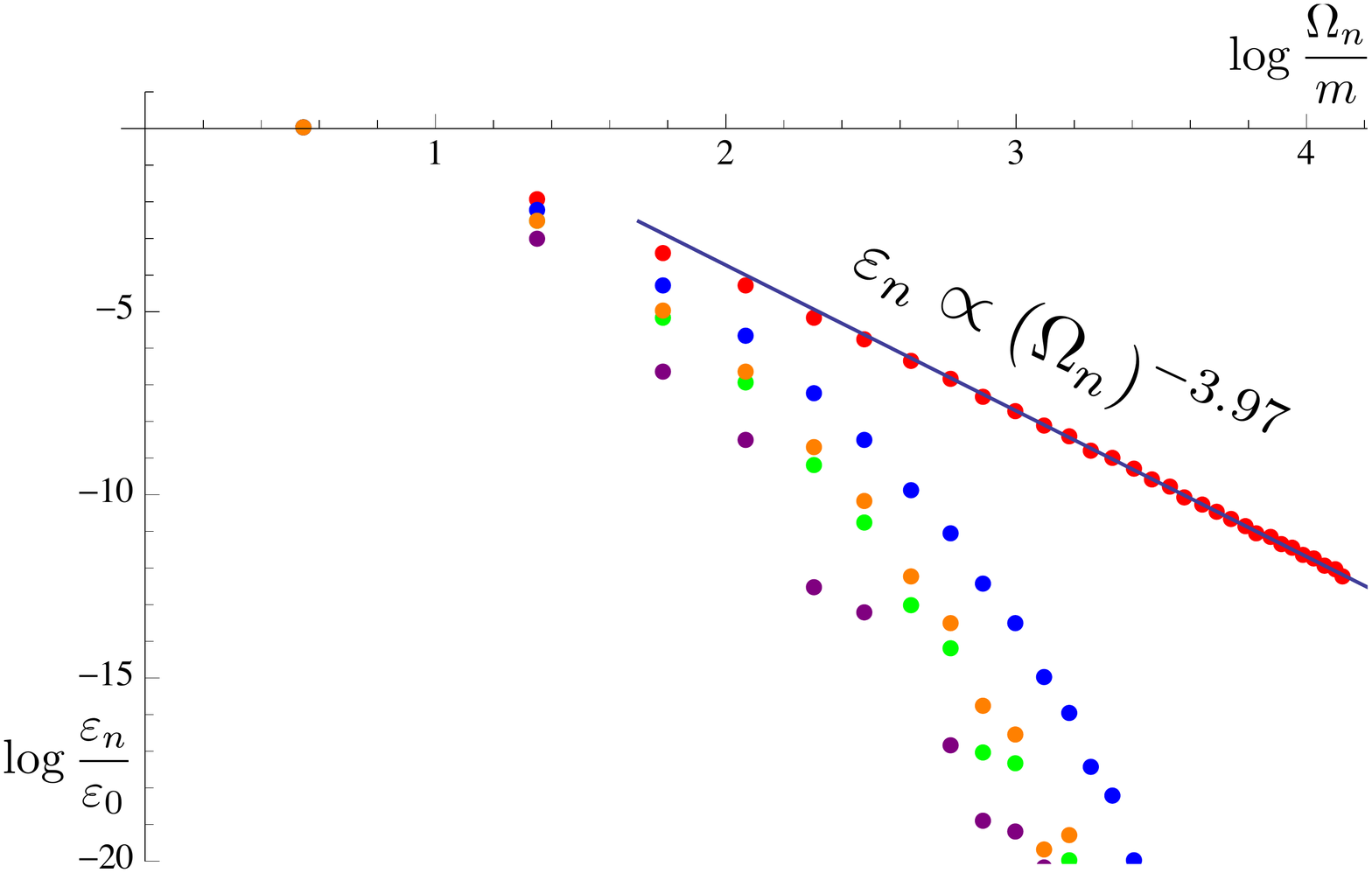}
    \caption{The power law of the $n$-th meson mass including a static electric field.  The vertical axis is the logarithm plots of the $n$-th meson energy divided by the lowest meson energy. The transverse axis describes the logarithm plots of the $n$-th meson mass. }
    \label{fig:mesonE}
  \end{center}
\end{figure}

From these numerical shape of the D5-branes in Fig.~\ref{fig:D5shapeE}, 
we can calculate the meson energy
spectrum, through the definitions given in the previous subsection.
As the shape of the D5-brane changes by varying $\mathcal{E}/\mathcal{E}_{\mathrm{cr}}$,
the decomposed meson energy spectrum changes. 
Our result for the energy distribution $\varepsilon_n$ is presented in Fig.~\ref{fig:mesonE}.

The energy distribution Fig.~\ref{fig:mesonE} shows that the critical embedding
(the red dots) is distinctively different from other Minkowski embeddings. 
The red dots can be linearly fit as 
\begin{align}
\varepsilon_n \propto(\Omega_n)^{-3.97}
\label{-3.97}
\end{align}
which is a power law. This is nothing but a weak turbulence, and is very similar to
what we has been known in \cite{Hashimoto:2014xta, Hashimoto:2014dda} and 
what we found in the previous section. For the other Minkowski embeddings,
there is a significant reduction of energy for higher excited mesons (large $n$ region).
So, the power law appears only at the critical embedding.
%
%
The cusp of the D5-brane seems to be responsible for the power law of the meson energy
distribution. 
We can conclude that 
the weak turbulence of the excited mesons is caused by the cusp of the D5-brane,
which is realized and accompanied with the phase transition.


\subsection{Turbulence at a finite temperature}

In this subsection, we consider the energy spectrum of the highly excited meson
states in a finite temperature D3-D5 system in the absence of the electric field,
to study the universality of the power law (\ref{-3.97}).
The basis of the analysis was given in  \cite{Mateos:2007vn} as we have reviewed.
In the previous subsection, we studied the turbulent 
meson spectrum in the electric field. When the probe D5-brane has a cusp, 
the meson energy distribution obeys a power law: $\varepsilon_n\propto(\Omega_n)^{-3.97}$. 
It is of interest if the power law is universal or not.
So, here we introduce a temperature to the AdS background, and examine 
the power law.

In AdS/CFT correspondence, the temperature is introduced by replacing the
background AdS geometry (\ref{AdS}) by an AdS black hole metric, 
\begin{align}
ds^{2} = \frac{r^2}{R^2}[-f(r)dt^2 + d\vec{x}^2] + \frac{R^2}{r^2}\left[\frac{dr^2}{f(r)} + r^2d\Omega^{2}_{5} \right] ,
\end{align}
where $d\vec{x}^2\equiv dx^2_1+dx^2_2+dx^2_3$, and 
the function of $f(r)$ is defined by $f(r) \equiv 1-(r_\mathrm{H}/r)^4 $. 
The location of the black hole horizon $r_\mathrm{H}$ is related 
to the temperature as $T=r_\mathrm{H}/\pi R^2$ due to the AdS/CFT dictionary. 
Using a new coordinate 
$2u^2 = r^2 + \sqrt{r^4 - r_\mathrm{H}^4}$,
the AdS black hole metric is written as
\begin{align}
ds^{2} = \frac{u^2}{R^2}\left[-\frac{f(u)^2}{\tilde{f}(u)}dt^2 + \tilde{f}(u)d\vec{x}^2\right] + \frac{R^2}{u^2}\left[dv^2 + v^2d\Omega^{2}_{2} + d\tilde{\omega}^2_4 + d\tilde{\omega}^2_5 + d\tilde{\omega}^2_6 \right],
\end{align}
where 
\begin{align}
f(u)\equiv 1-\frac{r_{\mathrm{H}}^4}{4u^4}, \hspace{4mm}\tilde{f}(u) \equiv 1+\frac{r_{\mathrm{H}}^4}{4u^4}\, .
\end{align}
The radial coordinates $u$ and $v$ parameterize the radii of $S^5$ and $S^2$ respectively,
and they are related as 
$u^2=v^2+\tilde{\omega}_4^2+\tilde{\omega}_5^2+\tilde{\omega}_6^2$. 
As before, using the rotational symmetry, we can restrict ourselves to
$\tilde{\omega}_5=\tilde{\omega}_6=0$
and 
$\tilde{\omega}_4\equiv \tilde{L}(v)$. Then, the induced metric on the D5-brane is given by
\begin{align}
ds^{2} = \frac{u^2}{R^2}\left[-\frac{f(u)^2}{\tilde{f}(u)}dt^2 + \tilde{f}(u)d\vec{x}^2\right] + \frac{R^2}{u^2}\left[\{1 + (\partial_v \tilde{L})^2\}dv^2 + v^2d\Omega^{2}_{2}\right] .\label{D5T}
\end{align}

\begin{figure}[htbp]
  \begin{center}
    \includegraphics[width=8.0cm,bb=0 0 842 595]{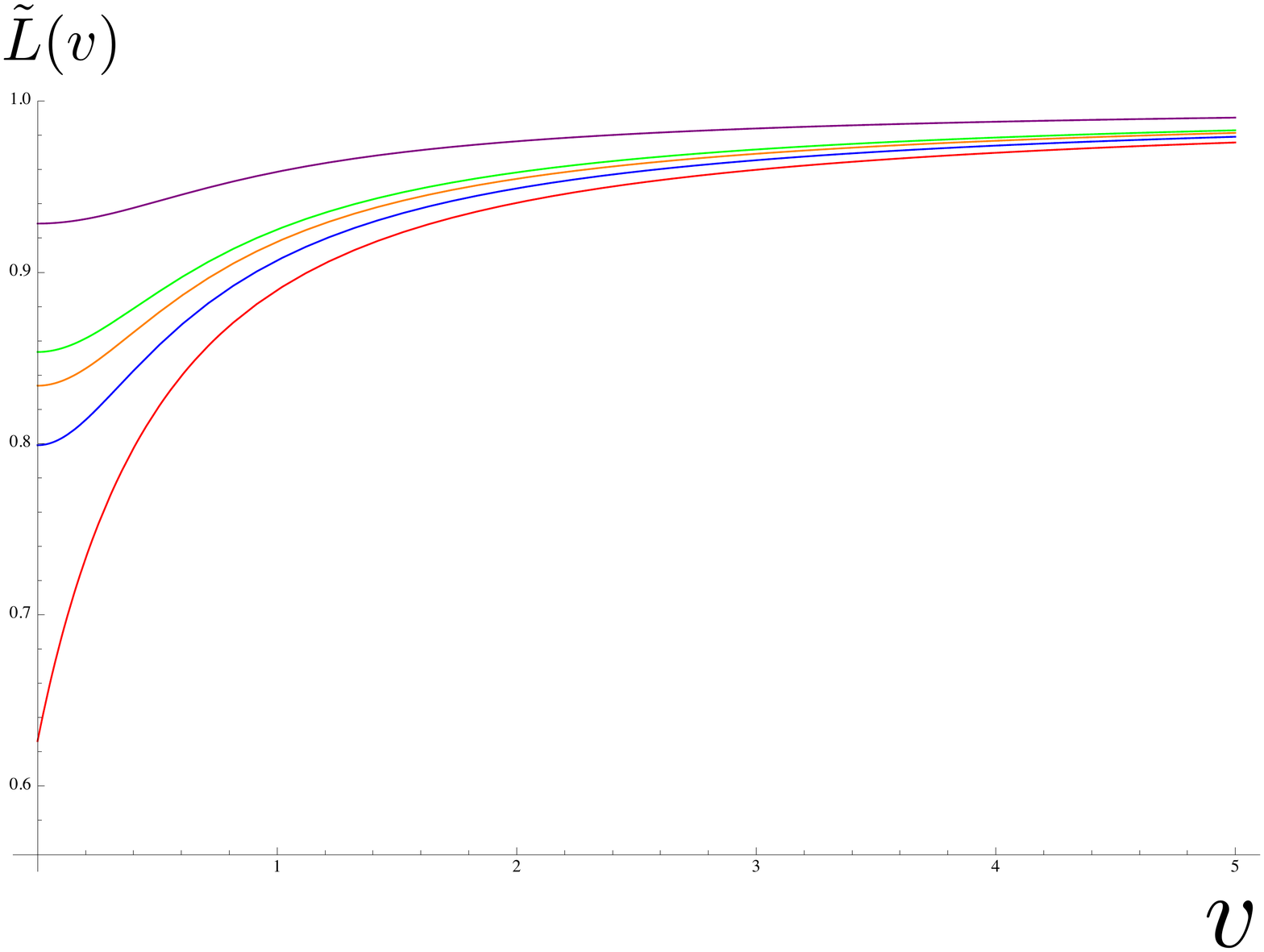}
    \caption{The shape of the D5-brane in the AdS black hole background. 
Each curve is with $r_{\mathrm{H}}/r_{\mathrm{cr}}=0.9,0.99,0.99,1.01,1$ respectively from top to bottom. 
The D5-brane of the red curve has a cusp at $\rho=0$, which we call a critical embedding, 
for 
$r_{\mathrm{H}}=r_{\mathrm{cr}}$$(\simeq 0.443)$. 
The other curves are Minkowski embeddings.}
    \label{fig:D5shapeT}
  \end{center}
\end{figure}

Next, we determine the shape 
$\tilde{L}(v)$ of the probe D5-brane in this finite temperature system. The 1-flavor DBI action is defined by (\ref{DBI}). By using the induced metric (\ref{D5T}), the DBI action becomes
\begin{align}
S_{\mathrm{DBI}} = - 4\pi\tau_5\int{d^3 x}\int_{0}^{\infty}\!\!\!\!\!{dv}\ v^2\left(1-\frac{r_{\mathrm{H}}^4}{4(v^2 + \tilde{L}^2)^2}\right)\sqrt{1+\frac{r_{\mathrm{H}}^4}{4(v^2 + \tilde{L}^2)^2}}\sqrt{1+(\partial_v \tilde{L})^2}.
\end{align}
The Euler-Lagrange equation of the D5-brane is obtained as
\begin{align}
\partial_v\left[\frac{v^2\partial_v\tilde{L}\left(4(v^2+\tilde{L}^2)^2 - r_{\mathrm{H}}^4\right) \sqrt{4(v^2+\tilde{L}^2)^2 + r_{\mathrm{H}}^4}}{(v^2+\tilde{L}^2)^3\sqrt{1+(\partial_v\tilde{L})^2}}\right] \hspace{20mm}\\ \notag 
 - \frac{2r_{\mathrm{H}}^4v^2\tilde{L}\left(4(v^2+\tilde{L}^2)^2 + 3r_{\mathrm{H}}^4\right)\sqrt{1+(\partial_v\tilde{L})^2}}{(v^2+\tilde{L}^2)^4\sqrt{4(v^2+\tilde{L}^2)^2 + r_{\mathrm{H}}^4}} = 0.
\end{align}
We numerically calculate classical solutions $\tilde{L}(v)$ of the equation of motion.
We chose the convention $R=1$ and $m=1$ which is the same as before.
The theory has only two scales, the quark mass and the temperature. So the theory
is determined only by the ratio of those. 
For fixed $m=1$, we vary $r_\mathrm{H}$ and obtain various D5-brane shapes, as shown 
in Fig.~\ref{fig:D5shapeT}.

As the temperature changes, the amount of the gravity which the D5-brane feels changes,
since the location of the horizon approaches the D5-brane.
The red curve in Fig.~\ref{fig:D5shapeT} has a cusp of the D5-brane,  and shows 
that the D5-brane is at the critical embedding.

\begin{figure}[htbp]
  \begin{center}
    \includegraphics[width=8.0cm,bb=0 0 842 595]{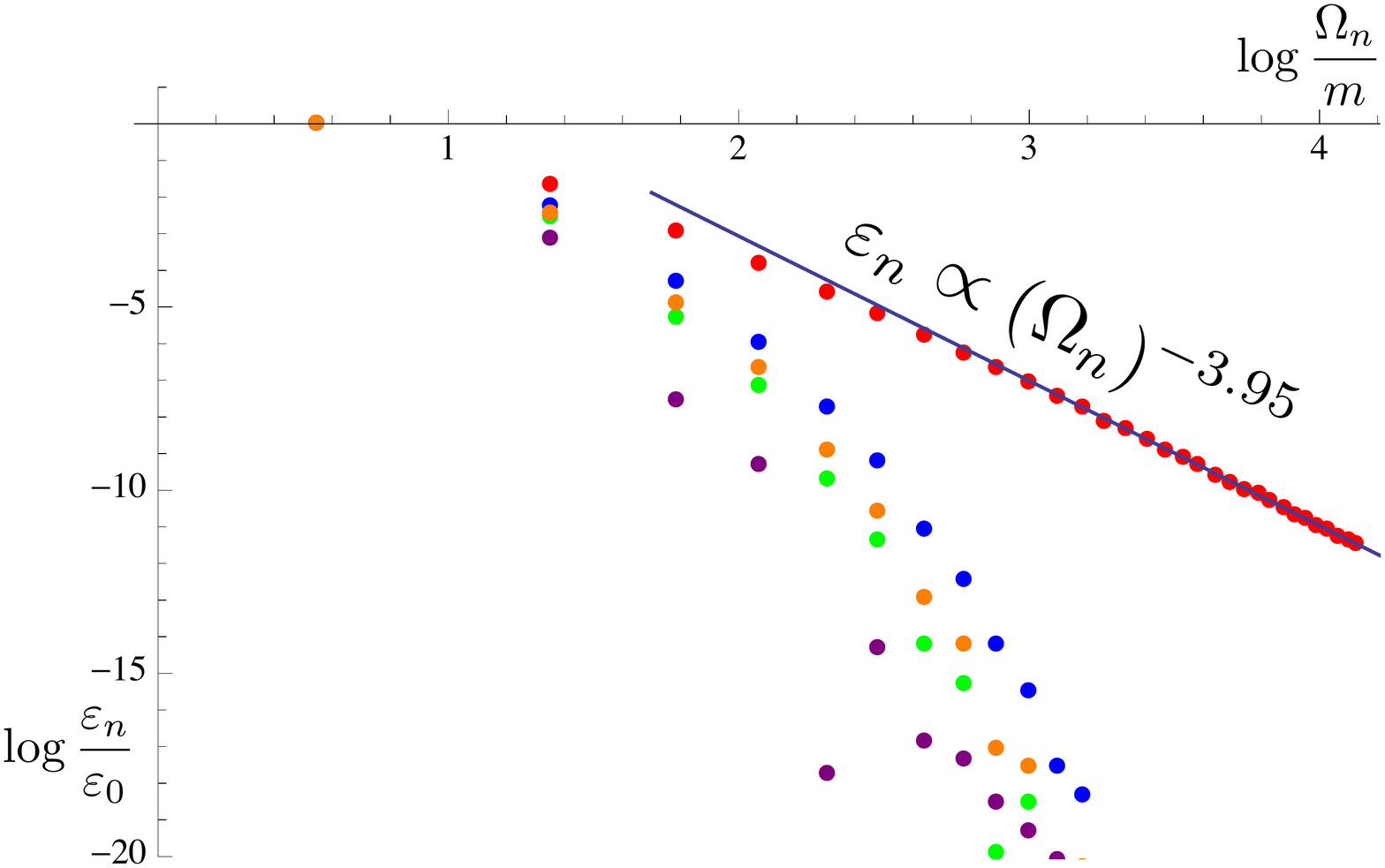}
    \caption{The power law of the energy distribution $\varepsilon_n$ for the $n$-th meson
    resonance, in the finite temperature system.  The vertical axis is the logarithm of the $n$-th meson energy $\varepsilon_n$ divided by the lowest meson energy $\varepsilon_0$. The horizontal axis is for  the logarithm of the $n$-th meson mass $\Omega_n$. The colors of the dots correspond to that of the curves in the previous figure.}
    \label{fig:mesonT}
  \end{center}
\end{figure}

We expect that a turbulent behavior of meson excited states causes 
the cusp of the D5-brane at $r_{\mathrm{H}}/r_{\mathrm{cr}}=1$ shown by the red curve
in Fig.~\ref{fig:D5shapeT}. 
So let us present the results of the meson energy spectrum for 
the higher excitation modes: see Fig.\ref{fig:mesonT}.
As expected, the red dots which are the energy distribution
of meson excitations corresponding to the cusp D5-brane (the red curve in Fig.~\ref{fig:D5shapeT}) has a linear behavior. A linear fit of the red dots shows
\begin{align}
\varepsilon_n\propto(\Omega_n)^{-3.95}. 
\end{align}
Again, we have found that the critical embedding of the D5-brane shows a turbulent
behavior of mesons. For other curves with the Minkowski embedding, we have no
power law: the energy deposit at the higher meson resonance decreases rapidly 
for large $n$. So the weak turbulence, the power law behavior, is unique to
the critical embedding of the D5-brane.

%

In conclusion, the turbulent behavior of the higher excited modes of mesons
is observed when the shape of D5-brane has a cusp in the static electric field or at the finite temperature. 
The meson energy distribution obeys the power law as a function of the meson mass, 
$\alpha\simeq -3.97$ or $\beta\simeq -3.95$ in the static electric field or at the finite temperature system, respectively. Since these powers are quite close to $-4$, we conclude
that 
the turbulent behavior is a universal phenomenon irrespective of how the transition is made. 
We expect that the exact power may be 
$\varepsilon_n\propto(\Omega_n)^{-4}$, which may be confirmed
with numerical simulations with higher accuracy.

\section{Universal turbulence and a  conjecture}

In the previous sections, we found that the power associated with the
level distribution $\varepsilon_n$ of the energy density for the meson level $n$
has an integer power: 
\begin{eqnarray}
\varepsilon_n \propto (m_n)^\alpha\, ,
\label{conj}
\end{eqnarray}
where $\alpha = -5$ for the D3/D7 system (treated in section 2) and
$\alpha=-4$ for the D3/D5 system (treated in section 3).
Furthermore, the power in each case is found to be universal. The power does not
change even though the external field to produce the phase transition is
modified --- the temperature and the magnetic field.

Therefore, it seems that the power $\alpha$ depends only on the dimensionality of the
brane cone, and it does not depend on how the phase transition is driven. 
It is natural to make a generic conjecture on the power $\alpha$: 
{\it 
The 
energy distribution for level $n$ at the conical brane configuration for the
phase transition is given by the power law as in (\ref{conj}), and the power
is determined as
}
\begin{eqnarray}
\alpha = -(d_{\rm cone}+1)
\label{conjd}
\end{eqnarray}
{\it 
where $d_{\rm cone}$ is the dimension of the cone.
}
For the case of the D3/D7 (D3/D5) system,
$d_{\rm cone}=4$ ($d_{\rm cone}=3$).

In this paper, we presented evidence for this conjecture for various 
situations associated with the D3/D7 and the D3/D5 holographic models,
with various external fields, and all are consistent with the above conjecture.

To strengthen the plausibility of the conjecture, in the following part of this section 
we provide simple examples of decomposition of a conic brane configuration
in flat space by eigen modes of harmonic functions. The examples are
1-dimensional and 2-dimensional cones, for simplicity. Both nontrivially
can be worked out and are consistent with the conjectured relation (\ref{conjd}).

\begin{description}

\item{\bf Ex. 1-dimensional cone}

Let us consider a one-dimensional cone in a flat space and calculate the power
$\alpha$ when it is expanded in eigen modes.
The one-dimensional cone is simply written as
\begin{eqnarray}
L(\rho) = 1- \frac{2}{\pi} |\rho| 
\end{eqnarray}
which is defined on an interval $-\pi/2 < \rho < \pi/2$, at whose boundary
the height function $L(\rho)$ is put to zero. We expand this cone by 
harmonic functions in one-dimensional flat space, {\it i.e.} Fourier modes.
The eigen basis satisfying the Dirichlet boundary condition at the boundaries
of the interval is
\begin{eqnarray}
e_n(\rho) = \sqrt{\frac{2}{\pi}} \cos \left((2n+1)\rho\right) 
\end{eqnarray}
where $n=0,1,2,\cdots$ specifies the level of the modes. The magnitude of
the eigenvalue
of the Laplacian $\partial_\rho^2$ is just $m_n^2 = (2n+1)^2$.
The expansion is easily done as
\begin{eqnarray}
L(\rho) = \sum_{n=0}^\infty c_n e_n(\rho) \, 
\quad
\mbox{where}
\quad
c_n \equiv \frac{2^{5/2}}{\pi^{3/2}}  \frac{1}{(2n+1)^2} \, . 
\end{eqnarray}
So the meson energy for each level $n$ is 
\begin{eqnarray}
\varepsilon_n \equiv \frac12 m_n^2 c_n^2
= \frac{2^4}{\pi^3} \frac{1}{(2n+1)^2}
\end{eqnarray}
So, for large $n$, we obtain the power
\begin{eqnarray}
\varepsilon_n \sim n^{\alpha} \, , \quad \alpha = -2 \, .
\end{eqnarray}
From this simple example, we find that the possible power in the 
meson melting transition, if occurs in holography, would have $\alpha=-2$,
if the cone of the probe brane is just one-dimensional.

\item{\bf Ex. 2-dimensional cone}

Let us proceed to an example in 2 dimensions, to support our conjecture.
The example is again a cone in a flat space. Let us consider a 2-dimensional cone
\begin{eqnarray}
L(\rho) = 1- \rho
\label{2dcone}
\end{eqnarray}
where $\rho \equiv \sqrt{(x^1)^2 + (x^2)^2}$, and we consider $0\leq \rho \leq 1$. 
The boundary $\rho=1$ is a circle, and we impose a Dirichlet boundary condition for
harmonic functions on the disk.
The harmonic functions in 2 dimensions are Bessel functions $J_\nu(r)$,
\begin{eqnarray}
\left[
\frac{d^2}{dr^2} + \frac{1}{r} \frac{d}{dr}
+ \left(1 - \frac{\nu^2}{r^2}\right)
\right] J_\nu(r)=0 \, .
\end{eqnarray}
The subscript $\nu$ corresponds to the angular mode, so our cone configuration
should be expanded only with $\nu=0$. The Dirichlet boundary condition
says $e_n(\rho=1)=0$, therefore, in terms of the Bessel functions, we have
\begin{eqnarray}
e_n(\rho) = \frac{\sqrt{2}}{|J_0'(r_n)|} J_0(r_n \rho)
\end{eqnarray}
where $r_n (n=0,1,2,3,\cdots)$ label zeros of the Bessel function $J_0(r)$.
The normalization is already fixed by using the following formula
\begin{eqnarray}
\int_0^1 \rho \left(J_0(r_n \rho)\right)^2 d\rho = \frac12 \left(J'_0(r_n)\right)^2
\, .
\end{eqnarray}
The eigenvalue of the eigen mode function $e_n(\rho)$ is $r_n^2$, thus the 
``meson mass" of level $n$ is $m_n = r_n$.

The expansion of the cone (\ref{2dcone}) is given by
\begin{eqnarray}
L(\rho) = \sum_{n=0}^\infty c_n e_n(\rho), 
\quad
c_n \equiv \int_0^1 \rho (1-\rho) e_n(\rho)\, .
\label{2exp}
\end{eqnarray}
We are interested in only the large $n$ behavior, so we use 
the following expression for the large $n$ asymptotic expansion of the Bessel functions,
\begin{eqnarray}
J_0(r) = \sqrt{\frac{2}{\pi r}}
\left(
\cos (r-\pi/4) + \frac{1}{8r} \sin (r-\pi/4) + {\cal O}(1/r^2)
\right) \, .
\end{eqnarray}
The zeros of this function at large $n$ are given by
\begin{eqnarray}
r_n \sim \frac{4n+3}{4}\pi + \frac{1}{(8n+6)\pi} \quad (n \gg 1)
\end{eqnarray}
So, at the leading order in the large $n$ expansion, we have
\begin{eqnarray}
e_n(\rho) \sim \sqrt{\frac{2}{\rho}} \cos \left(\frac{(4n+3)\pi}{4} \rho - \frac{\pi}{4}\right)  \, .
\end{eqnarray}
Substituting this into (\ref{2exp}), 
we obtain the large $n$ behavior of the coefficient $c_n$, 
\begin{eqnarray}
c_n = \frac{3\sqrt{2}}{4\pi^2} n^{-5/2} + {\cal O}(n^{-3}) \, .
\end{eqnarray}
The meson condensation $c_n$ is found to scale as $\sim n^{-5/2}$. 
Therefore the energy distribution is obtained as
\begin{eqnarray}
\varepsilon_n = \frac{1}{2} m_n^2 c_n^2 \sim 
\frac{9}{16\pi^2} n^{-3}  \, ,
\end{eqnarray}
which shows the power  
\begin{eqnarray}
\alpha=-3
\end{eqnarray}
for the case of the 2-dimensional cone in a flat space.
\end{description}

These simple examples are consistent with our conjecture (\ref{conjd}),
so it is expected that the brane turbulence (\ref{conj}) with the power (\ref{conjd})
is universal, not only in AdS-like geometries but also in a box of flat geometries.


\section{Conclusions and discussions}

In this paper, we examined the universality of the power $\alpha$
which appears in the meson turbulence relation (\ref{turbulent2}).
The turbulent meson behavior was discovered in \cite{Hashimoto:2014xta,Hashimoto:2014dda} for
a holographic system made by D3-D7 configurations which is
dual to ${\cal N}=2$ supersymmetric QCD at large $N_c$ and 
strong coupling limit. 
The power $\alpha=-5$ was found in \cite{Hashimoto:2014xta,Hashimoto:2014dda}
when the meson melting transition was driven by an external
electric field, either static or dynamical. 
We found that the power $\alpha$ is universal, by calculating the power
for the meson melting transition with magnetic fields in addition to the 
electric field, or with a finite temperature. For any transition we considered,
we found 
$\alpha=-5$ for the D3-D7 holographic system (Section 2), and $\alpha=-4$ for 
the D3-D5 system (Section 3).

Based on these findings with numerical calculations of $\alpha$, 
we made a conjecture $\alpha = -(d_{\rm cone} + 1)$ where $d_{\rm cone}$
is the number of dimensions of the cone formed by the probe D-brane.
For the probe D7-brane (D5-brane), the cone dimension is 
$d_{\rm cone}=4 (3)$, respectively.

Although we worked only 
with near horizon geometries produced by $N_c$ D3-branes,
we believe that other geometries with various other types of probe D-branes
in holography 
may exhibit the same turbulent relation (\ref{turbulent2}) with our universal power
$\alpha = -(d_{\rm cone} + 1)$ determined solely by the dimensions $d_{\rm cone}$
of the cone at the critical embedding.
For example, it should be interesting to work out the power numerically for
other geometries such as the ones made by D$p$-branes ($p\neq 3$) 
or M-branes.\footnote{While writing this manuscript, 
we noticed that a recent paper \cite{Ishii:2015wua} treats 
kinks on a fundamental string in a holographic setup. It may correspond to our
$d_{\rm cone}=1$ treated in Section 4, and the authors of 
\cite{Ishii:2015wua} found numerically  $\alpha\simeq -2$ for a transverse circular quench, which is consistent with our formula. However, other quenches show
different powers, so a further investigation for the universality may be necessary.}

The weak turbulence in holographic spacetime may be a quite generic phenomenon.
In this paper we analyzed the turbulent energy distribution on probe D-branes,
and found an intriguing universality. 
If the universality is applied to {\it any} probe sector of {\it any} holographic geometry,
then we can make a generic statement that at ``meson" melting transition
in holographic field theories the meson energy distribution obeys a turbulent
energy distribution (\ref{turbulent2}) with a power universally determined
by the dimensions of the probe conic brane. The dimensions apparently 
depends on the spatial dimensions of the internal space in the gravity dual.
A radical conjecture which follows from this argument is that the large $N_c$ QCD
at strong coupling limit at a quark-hadron transition 
may have the meson turbulence with the power $\alpha=-1$,
since QCD does not have any adjoint scalar field probing the internal space dimensions and thus it is expected to correspond to $d_{\rm cone}=0$.

The power $\alpha$ in the Kolmogorov-like power scaling of the turbulent relation
(\ref{turbulent2}) may be related to how easily the system is driven to the meson melting transition. Once some external field which drives the system to the melted meson phase is turned on, the system becomes unstable and starts to evolve dynamically. Since the instability under strong electromagnetic field, as well as
the time evolution, have been analyzed in holographic set-ups with D-brane effective actions (see for example 
\cite{Hashimoto:2013mua,Hashimoto:2014dza,Hashimoto:2014yza,Hashimoto:2014yya}),
it would be interesting to investigate possible relations to those. Relatedly, 
it is important to ask 
what kind of physical properties in phase diagrams of strongly coupled 
gauge theories is characterized by the Kolmogorov-like power $\alpha$ 
of the meson turbulence.


\acknowledgments
A.S.~would like to thank R.~Meyer for valuable comments, and 
K.H.~would like to thank S.~Kinoshita, K.~Murata and T.~Oka for useful discussions.
The work of K.~H.~was supported in part by JSPS KAKENHI Grant Numbers 15H03658, 15K13483.
The work of M.N.~was supported in part by JSPS Research Fellowship for Young Scientists.



\begin{thebibliography}{10}

     
\bibitem{Bizon:2011gg} 
  P.~Bizon and A.~Rostworowski,
  ``On weakly turbulent instability of anti-de Sitter space,''
  Phys.\ Rev.\ Lett.\  {\bf 107}, 031102 (2011)
  [arXiv:1104.3702 [gr-qc]].
  

\bibitem{Maldacena:1997re} 
  J.~M.~Maldacena,
  ``The Large N limit of superconformal field theories and supergravity,''
  Adv.\ Theor.\ Math.\ Phys.\  {\bf 2}, 231 (1998)
  [hep-th/9711200].

\bibitem{Gubser:1998bc}
 S.~Gubser, I.~R. Klebanov, and A.~M. Polyakov, 
 ``Gauge theory correlators from noncritical string theory,''  
{\em Phys.Lett.} B {\bf 428},105 (1998).
 
\bibitem{Witten:1998qj}
 E.~Witten, 
 ``Anti-de Sitter space and holography,''  
{\em   Adv.Theor.Math.Phys.} {\bf 2}, 253 (1998).


\bibitem{deOliveira:2012dt} 
  H.~P.~de Oliveira, L.~A.~Pando Zayas and E.~L.~Rodrigues,
  ``A Kolmogorov-Zakharov Spectrum in AdS Gravitational Collapse,''
  Phys.\ Rev.\ Lett.\  {\bf 111}, no. 5, 051101 (2013)
  [arXiv:1209.2369 [hep-th]].
\bibitem{Liebling:2012gv} 
  S.~L.~Liebling,
  ``Nonlinear collapse in the semilinear wave equation in AdS space,''
  Phys.\ Rev.\ D {\bf 87}, no. 8, 081501 (2013)
  [arXiv:1212.6970 [gr-qc]].
\bibitem{Dias:2012tq} 
  O.~J.~C.~Dias, G.~T.~Horowitz, D.~Marolf and J.~E.~Santos,
  ``On the Nonlinear Stability of Asymptotically Anti-de Sitter Solutions,''
  Class.\ Quant.\ Grav.\  {\bf 29}, 235019 (2012)
  [arXiv:1208.5772 [gr-qc]].
\bibitem{Maliborski:2012gx} 
  M.~Maliborski,
  ``Instability of Flat Space Enclosed in a Cavity,''
  Phys.\ Rev.\ Lett.\  {\bf 109}, 221101 (2012)
  [arXiv:1208.2934 [gr-qc]].
\bibitem{Buchel:2012uh} 
  A.~Buchel, L.~Lehner and S.~L.~Liebling,
  ``Scalar Collapse in AdS,''
  Phys.\ Rev.\ D {\bf 86}, 123011 (2012)
  [arXiv:1210.0890 [gr-qc]].
\bibitem{Buchel:2013uba} 
  A.~Buchel, S.~L.~Liebling and L.~Lehner,
  ``Boson stars in AdS spacetime,''
  Phys.\ Rev.\ D {\bf 87}, no. 12, 123006 (2013)
  [arXiv:1304.4166 [gr-qc]].
\bibitem{Bizon:2013gxa} 
  P.~Bizo?,
  ``Is AdS stable?,''
  Gen.\ Rel.\ Grav.\  {\bf 46}, no. 5, 1724 (2014)
  [arXiv:1312.5544 [gr-qc]].
\bibitem{Maliborski:2013via} 
  M.~Maliborski and A.~Rostworowski,
  ``Lecture Notes on Turbulent Instability of Anti-de Sitter Spacetime,''
  Int.\ J.\ Mod.\ Phys.\ A {\bf 28}, 1340020 (2013)
  [arXiv:1308.1235 [gr-qc]].
\bibitem{Maliborski:2013jca} 
  M.~Maliborski and A.~Rostworowski,
  ``Time-Periodic Solutions in an Einstein AdS?Massless-Scalar-Field System,''
  Phys.\ Rev.\ Lett.\  {\bf 111}, 051102 (2013)
  [arXiv:1303.3186 [gr-qc]].
\bibitem{Balasubramanian:2014cja} 
  V.~Balasubramanian, A.~Buchel, S.~R.~Green, L.~Lehner and S.~L.~Liebling,
  ``Holographic Thermalization, Stability of Anti?de Sitter Space, and the Fermi-Pasta-Ulam Paradox,''
  Phys.\ Rev.\ Lett.\  {\bf 113}, no. 7, 071601 (2014)
  [arXiv:1403.6471 [hep-th]].
\bibitem{Maliborski:2014rma} 
  M.~Maliborski and A.~Rostworowski,
  ``What drives AdS spacetime unstable?,''
  Phys.\ Rev.\ D {\bf 89}, no. 12, 124006 (2014)
  [arXiv:1403.5434 [gr-qc]].
\bibitem{Craps:2014vaa} 
  B.~Craps, O.~Evnin and J.~Vanhoof,
  ``Renormalization group, secular term resummation and AdS (in)stability,''
  JHEP {\bf 1410}, 48 (2014)
  [arXiv:1407.6273 [gr-qc]].
\bibitem{Horowitz:2014hja} 
  G.~T.~Horowitz and J.~E.~Santos,
  ``Geons and the Instability of Anti-de Sitter Spacetime,''
  arXiv:1408.5906 [gr-qc].
\bibitem{Dias:2011ss} 
  O.~J.~C.~Dias, G.~T.~Horowitz and J.~E.~Santos,
  ``Gravitational Turbulent Instability of Anti-de Sitter Space,''
  Class.\ Quant.\ Grav.\  {\bf 29}, 194002 (2012)
  [arXiv:1109.1825 [hep-th]].
\bibitem{Dimitrakopoulos:2014ada} 
  F.~V.~Dimitrakopoulos, B.~Freivogel, M.~Lippert and I.~S.~Yang,
  ``Instability corners in AdS space,''
  arXiv:1410.1880 [hep-th].
\bibitem{Caceres:2014pda} 
  E.~Caceres, A.~Kundu, J.~F.~Pedraza and D.~L.~Yang,
  ``Weak Field Collapse in AdS: Introducing a Charge Density,''
  arXiv:1411.1744 [hep-th].
\bibitem{Craps:2014jwa} 
  B.~Craps, O.~Evnin and J.~Vanhoof,
  ``Renormalization, averaging, conservation laws and AdS (in)stability,''
  JHEP {\bf 1501}, 108 (2015)
  [arXiv:1412.3249 [gr-qc]].
\bibitem{Buchel:2014xwa} 
  A.~Buchel, S.~R.~Green, L.~Lehner and S.~L.~Liebling,
  ``Conserved quantities and dual turbulent cascades in anti?de Sitter spacetime,''
  Phys.\ Rev.\ D {\bf 91}, no. 6, 064026 (2015)
  [arXiv:1412.4761 [gr-qc]].
  
      


\bibitem{Dias:2011ss}
  O.~J.~C.~Dias, G.~T.~Horowitz and J.~E.~Santos,
  ``Gravitational Turbulent Instability of Anti-de Sitter Space,''
  Class.\ Quant.\ Grav.\  {\bf 29} (2012) 194002
  [arXiv:1109.1825 [hep-th]].
\bibitem{Heller:2012je}
  M.~P.~Heller, R.~A.~Janik and P.~Witaszczyk,
  ``A numerical relativity approach to the initial value problem in asymptotically Anti-de Sitter spacetime for plasma thermalization - an ADM formulation,''
  Phys.\ Rev.\ D {\bf 85} (2012) 126002
  [arXiv:1203.0755 [hep-th]].
\bibitem{Dias:2012tq}
  O.~J.~C.~Dias, G.~T.~Horowitz, D.~Marolf and J.~E.~Santos,
  ``On the Nonlinear Stability of Asymptotically Anti-de Sitter Solutions,''
  Class.\ Quant.\ Grav.\  {\bf 29} (2012) 235019
  [arXiv:1208.5772 [gr-qc]].
\bibitem{Garfinkle:2011tc}
  D.~Garfinkle, L.~A.~Pando Zayas and D.~Reichmann,
  ``On Field Theory Thermalization from Gravitational Collapse,''
  JHEP {\bf 1202} (2012) 119
  [arXiv:1110.5823 [hep-th]].
\bibitem{Heller:2012km}
  M.~P.~Heller, D.~Mateos, W.~van der Schee and D.~Trancanelli,
  Phys.\ Rev.\ Lett.\  {\bf 108} (2012) 191601
  [arXiv:1202.0981 [hep-th]].
\bibitem{Hawking:2014tga}
  S.~W.~Hawking,
  ``Information Preservation and Weather Forecasting for Black Holes,''
  arXiv:1401.5761 [hep-th].
\bibitem{Buchel:2013uba}
  A.~Buchel, S.~L.~Liebling and L.~Lehner,
  ``Boson stars in AdS spacetime,''
  Phys.\ Rev.\ D {\bf 87} (2013) 12,  123006
  [arXiv:1304.4166 [gr-qc]].
\bibitem{Bantilan:2012vu}
  H.~Bantilan, F.~Pretorius and S.~S.~Gubser,
  ``Simulation of Asymptotically AdS5 Spacetimes with a Generalized Harmonic Evolution Scheme,''
  Phys.\ Rev.\ D {\bf 85} (2012) 084038
  [arXiv:1201.2132 [hep-th]].
\bibitem{Cardoso:2012qm}
  V.~Cardoso, L.~Gualtieri, C.~Herdeiro, U.~Sperhake, P.~M.~Chesler, L.~Lehner, S.~C.~Park and H.~S.~Reall {\it et al.},
  ``NR/HEP: roadmap for the future,''
  Class.\ Quant.\ Grav.\  {\bf 29} (2012) 244001
  [arXiv:1201.5118 [hep-th]].
\bibitem{Jalmuzna:2011qw}
  J.~Jalmuzna, A.~Rostworowski and P.~Bizon,
  ``A Comment on AdS collapse of a scalar field in higher dimensions,''
  Phys.\ Rev.\ D {\bf 84} (2011) 085021
  [arXiv:1108.4539 [gr-qc]].
\bibitem{Bhaseen:2012gg}
  M.~J.~Bhaseen, J.~P.~Gauntlett, B.~D.~Simons, J.~Sonner and T.~Wiseman,
  ``Holographic Superfluids and the Dynamics of Symmetry Breaking,''
  Phys.\ Rev.\ Lett.\  {\bf 110} (2013) 1,  015301
  [arXiv:1207.4194 [hep-th]].
\bibitem{Liebling:2012fv}
  S.~L.~Liebling and C.~Palenzuela,
  ``Dynamical Boson Stars,''
  Living Rev.\ Rel.\  {\bf 15} (2012) 6
  [arXiv:1202.5809 [gr-qc]].
\bibitem{GarciaBellido:2011de}
  J.~Garcia-Bellido, J.~Rubio, M.~Shaposhnikov and D.~Zenhausern,
  ``Higgs-Dilaton Cosmology: From the Early to the Late Universe,''
  Phys.\ Rev.\ D {\bf 84} (2011) 123504
  [arXiv:1107.2163 [hep-ph]].
\bibitem{Maliborski:2013jca}
  M.~Maliborski and A.~Rostworowski,
  ``Time-Periodic Solutions in an Einstein AdS?Massless-Scalar-Field System,''
  Phys.\ Rev.\ Lett.\  {\bf 111} (2013) 051102
  [arXiv:1303.3186 [gr-qc]].
\bibitem{Buchel:2012uh}
  A.~Buchel, L.~Lehner and S.~L.~Liebling,
  ``Scalar Collapse in AdS,''
  Phys.\ Rev.\ D {\bf 86} (2012) 123011
  [arXiv:1210.0890 [gr-qc]].
\bibitem{Kunduri:2013gce}
  H.~K.~Kunduri and J.~Lucietti,
  ``Classification of near-horizon geometries of extremal black holes,''
  Living Rev.\ Rel.\  {\bf 16} (2013) 8
  [arXiv:1306.2517 [hep-th]].
\bibitem{Wu:2012rib}
  B.~Wu,
  ``On holographic thermalization and gravitational collapse of massless scalar fields,''
  JHEP {\bf 1210} (2012) 133
  [arXiv:1208.1393 [hep-th]].


\bibitem{Karch:2002sh} 
  A.~Karch and E.~Katz,
  ``Adding flavor to AdS / CFT,''
  JHEP {\bf 0206}, 043 (2002)
  [hep-th/0205236].

\bibitem{Kruczenski:2003be} 
  M.~Kruczenski, D.~Mateos, R.~C.~Myers and D.~J.~Winters,
  ``Meson spectroscopy in AdS / CFT with flavor,''
  JHEP {\bf 0307}, 049 (2003)
  [hep-th/0304032].

\bibitem{Kruczenski:2003uq} 
  M.~Kruczenski, D.~Mateos, R.~C.~Myers and D.~J.~Winters,
  ``Towards a holographic dual of large N(c) QCD,''
  JHEP {\bf 0405}, 041 (2004)
  [hep-th/0311270].

\bibitem{Sakai:2004cn} 
  T.~Sakai and S.~Sugimoto,
  ``Low energy hadron physics in holographic QCD,''
  Prog.\ Theor.\ Phys.\  {\bf 113}, 843 (2005)
  [hep-th/0412141].
  
\bibitem{Erdmenger:2007cm} 
  J.~Erdmenger, N.~Evans, I.~Kirsch and E.~Threlfall,
  ``Mesons in Gauge/Gravity Duals - A Review,''
  Eur.\ Phys.\ J.\ A {\bf 35}, 81 (2008)
  [arXiv:0711.4467 [hep-th]].




\bibitem{Hashimoto:2014xta} 
  K.~Hashimoto, S.~Kinoshita, K.~Murata and T.~Oka,
  ``Turbulent meson condensation in quark deconfinement,''
  arXiv:1408.6293 [hep-th].
  
\bibitem{Hashimoto:2014dda} 
  K.~Hashimoto, S.~Kinoshita, K.~Murata and T.~Oka,
  ``Meson turbulence at quark deconfinement from AdS/CFT,''
  arXiv:1412.4964 [hep-th].

\bibitem{Hashimoto:2014yza} 
  K.~Hashimoto, S.~Kinoshita, K.~Murata and T.~Oka,
  ``Electric Field Quench in AdS/CFT,''
  JHEP {\bf 1409}, 126 (2014)
  [arXiv:1407.0798 [hep-th]].

\bibitem{Ishii:2014paa} 
  T.~Ishii, S.~Kinoshita, K.~Murata and N.~Tanahashi,
  ``Dynamical Meson Melting in Holography,''
  JHEP {\bf 1404}, 099 (2014)
  [arXiv:1401.5106 [hep-th]].


\bibitem{Erdmenger:2007bn} 
  J.~Erdmenger, R.~Meyer and J.~P.~Shock,
  ``AdS/CFT with flavour in electric and magnetic Kalb-Ramond fields,''
  JHEP {\bf 0712}, 091 (2007)
  [arXiv:0709.1551 [hep-th]].

\bibitem{Albash:2007bq} 
  T.~Albash, V.~G.~Filev, C.~V.~Johnson and A.~Kundu,
  ``Quarks in an external electric field in finite temperature large N gauge theory,''
  JHEP {\bf 0808}, 092 (2008)
  [arXiv:0709.1554 [hep-th]].

\bibitem{Mateos:2006nu} 
  D.~Mateos, R.~C.~Myers and R.~M.~Thomson,
  ``Holographic phase transitions with fundamental matter,''
  Phys.\ Rev.\ Lett.\  {\bf 97}, 091601 (2006)
  [hep-th/0605046].
\bibitem{Frolov:2006tc} 
  V.~P.~Frolov,
  ``Merger Transitions in Brane-Black-Hole Systems: Criticality, Scaling, and Self-Similarity,''
  Phys.\ Rev.\ D {\bf 74}, 044006 (2006)
  [gr-qc/0604114].






\bibitem{Myers:2006qr} 
  R.~C.~Myers and R.~M.~Thomson,
  ``Holographic mesons in various dimensions,''
  JHEP {\bf 0609}, 066 (2006)
  [hep-th/0605017].

\bibitem{Mateos:2007vn} 
  D.~Mateos, R.~C.~Myers and R.~M.~Thomson,
  ``Thermodynamics of the brane,''
  JHEP {\bf 0705}, 067 (2007)
  [hep-th/0701132].
  




\bibitem{Ishii:2015wua} 
  T.~Ishii and K.~Murata,
  ``Turbulent strings in AdS/CFT,''
  arXiv:1504.02190 [hep-th].

\bibitem{Hashimoto:2013mua} 
  K.~Hashimoto and T.~Oka,
  ``Vacuum Instability in Electric Fields via AdS/CFT: Euler-Heisenberg Lagrangian and Planckian Thermalization,''
  JHEP {\bf 1310}, 116 (2013)
  [arXiv:1307.7423].

\bibitem{Hashimoto:2014dza} 
  K.~Hashimoto, T.~Oka and A.~Sonoda,
  ``Magnetic instability in AdS/CFT: Schwinger effect and Euler-Heisenberg Lagrangian of supersymmetric QCD,''
  JHEP {\bf 1406}, 085 (2014)
  [arXiv:1403.6336 [hep-th]].

\bibitem{Hashimoto:2014yya} 
  K.~Hashimoto, T.~Oka and A.~Sonoda,
  ``Electromagnetic instability in holographic QCD,''
  arXiv:1412.4254 [hep-th].
  
\end{thebibliography}
\end{document}